\documentclass[letterpaper, 10 pt, conference]{ieeeconf}  % Comment this line out if you need a4paper

\usepackage{enumerate}% http://ctan.org/pkg/enumerate
\usepackage{listings}
\usepackage{graphicx}
\usepackage{biblatex}
\usepackage{subfigure}
\usepackage{balance}
\usepackage{hyperref, graphicx, url, pslatex, paralist}

\title{Location, Location, Location: Exploring Amazon EC2 Spot Instance Pricing Across Geographical Regions - Extended Version
}

\author{Nnamdi Ekwe-Ekwe and Adam Barker\\
School of Computer Science, University of St Andrews, UK
\\ Email: \texttt{\{nnee, adam.barker\}@st-andrews.ac.uk}
}

\begin{document}

\maketitle

%%%%%%%%%%%%%%%%%%%%%%%%%%%%%%%%%%%%%%%%%%%%%%%%%%%%%%%%%%%%%%%%%%%%%%%%%%%%%%%%
\begin{abstract}

Cloud computing is becoming an almost ubiquitous part of the computing landscape. For many companies today, moving their entire infrastructure and workloads to the cloud reduces complexity, time to deployment, and saves money. Spot Instances, a subset of Amazon's cloud computing infrastructure (EC2), expands on this. They allow a user to bid on spare compute capacity in Amazon's data centres at heavily discounted prices. If demand was ever to increase such that the user's maximum bid is exceeded, their instance is terminated. 

In this paper, we conduct one of the first detailed analyses of how location affects the overall cost of deployment of a spot instance. We analyse pricing data across all available Amazon Web Services regions for 60 days for a variety of spot instance types. We relate the data we find to the overall AWS region as well as to the Availability Zone within that region.

We conclude that location does play a critical role in spot instance pricing and also that pricing differs depending on the granularity of that location - from a more coarse-grained AWS region to a more fine-grained Availability Zone within a region. We relate the pricing differences we find to the price's reliability, 
confirming whether one can be confident in the prices reported and subsequently, in the ensuing bids one makes.

We conclude by showing that it is possible to run workloads on Spot Instances achieving both a very low risk of termination as well as paying very low amounts per hour.
\end{abstract}

%%%%%%%%%%%%%%%%%%%%%%%%%%%%%%%%%%%%%%%%%%%%%%%%%%%%%%%%%%%%%%%%%%%%%%%%%%%%%%%%
\section{INTRODUCTION}

Amazon EC2 allows developers to provision compute resources, configure them to their needs, as well as scale resources up or down depending on application requirements. EC2 is widely used by organisations and developers as it makes it easy for them to provision compute resources on-demand without having to invest considerable time and money into building the underlying infrastructure needed in a data center. 

With EC2, developers have three major ways to pay what they provision. Developers can pay ``On-Demand" which allows the developer to pay for what they use by the hour with no ``long-term commitments" \cite{amazonec2pricingwebsite}. Developers can also pay via a ``Reserved Instance" model allowing them to commit to using a set of instances for either a ``1 or 3 year term" \cite{amazonec2pricingwebsite}. Amazon claims that this model provides a ``significant discount (up to 75\%) as compared to On-Demand pricing" \cite{amazonec2pricingwebsite}. Finally, developers can pay for resources using a ``Spot Instance" model \cite{amazonec2pricingwebsite}. The ``Spot Instance" model allows developers to bid on spare compute capacity giving them savings of up to ``90\% off the On-Demand price" \cite{amazonec2pricingwebsite}. 

Spot Instances are becoming increasingly useful for certain use cases. For example, users that suddenly have ``urgent computing needs for large amounts of additional capacity" \cite{amazonec2pricingwebsite} would find Spot Instances very useful. The user has access to both a sizeable amount of compute resources \textit{and} at very low prices. The only caveat with this approach, however, is that the application that the user is running must be fault-tolerant due to the potential of an instance being terminated at any time.

The majority of research into Spot Instances has focused on modelling their pricing strategy as well as determining the best bid prices to make on an instance \cite{artur-andrzejak, ben-yehuda, liang-zheng}. Papers such as \cite{artur-andrzejak, liang-zheng} have gone further - looking at how to find the lowest possible price for a user to spend, whilst maintaining the highest availability possible for the instance.

There is a distinct lack of research, however, which considers \emph{where} (in terms of region and availability zone) a user can \emph{reliably} deploy a spot instance in order to \emph{minimise cost}. In order to address this gap, we make a number of core research contributions: 

\begin{itemize}

\item We conduct one of the first detailed analyses of how location affects the overall cost of deployment of a spot instance. 

\item We analyse pricing data across all available Amazon Web Services regions for a variety of spot instance types. We relate the data we find to the overall AWS region as well as to the Availability Zone (AZ) within that region.

\item For any pricing differences we find, we check whether those differences are \emph{reliable} and as a result whether we can be confident in the ensuing bids we make.

\end{itemize}

We conclude that, 1) granularity of location has a significant impact on the overall price/price reliability of that instance; 2) the power of the instance type does not have as direct of an impact on price; some regions are universally substantially cheaper for certain instance types than others, while others are substantially more expensive.

The rest of this paper is structured as follows. Section II of this paper will examine what data we obtained from Amazon and its underlying structure. 
Section III will focus on the types of analyses we ran in order to explore the data. Section IV will give the results of our analyses. Section V will perform cross comparisons on our results to glean any interesting insights. Section VI will discuss related work.  Finally Section VII will summarise and give our overall conclusions as well as discuss our future direction and work going forward.

\section{EC2 SPOT PRICING DATA}

Amazon provides spot price data to any Amazon user for a period of up to 90 days from when a request is made. This is done via its API endpoint \cite{amazonspotinstanceapiwebsite}. This price data comes in JSON format in the form of a price point per time period.

For example:

\begin{verbatim}
 {
  "Timestamp": "2017-06-25T00:20:56.000Z",
  "ProductDescription": "Linux/UNIX",
  "InstanceType": "d2.2xlarge",
  "SpotPrice": "0.177800",
  "AvailabilityZone": "eu-west-2a"
 }
\end{verbatim}

In the above example, the instance type (\emph{d2.2xlarge}) was \$0.177800 (per hour) at 00:20:56 on the 25th of June 2017. The instance type was a Linux/UNIX instance - shown in the  ``ProductDescription" key. Finally, the Availability Zone in the EC2 region is \emph{eu-west-2a}. 
We do not conduct any research related to the ``ProductDescription" and so omit this column before we begin our analysis.

We used a data timeline of 60 days for our analysis. At time of carrying out this research, there were issues with retrieving the full 90 days worth of data from some zones. As a result, we worked with only 60 days worth of information. Additionally, not all Amazon regions provide a Spot Instance capability so our dataset comprised of data from only 4 AWS Regions - the EU, US, Asia Pacific and Canada Regions. 
These Amazon regions comprised of what, for the purposes of this paper, we'll call \textit{sub-regions}: EU - (\textit{eu-central-1, eu-west-1, eu-west-2}), US (\textit{us-east-1, us-east-2, us-west-1, us-west-2}), AP (\textit{ap-southeast-1, ap-southeast-2}) and finally, Canada (\textit{ca-central-1}). At time of running the analysis, there was a problem retrieving data from the \textit{ap-northeast-1} \emph{sub-region} and so those results have been omitted from the dataset.

\begin{table}[!htb]
    \centering
    \begin{tabular}{ |c|c|c|c| } 
        \hline
        Instance Type & Mean$\pm$Standard Deviation(\$p/h)\\
        \hline
        \textit{c3.large} & 0.077$\pm$0.077 \\
        \textit{c4.large} & 0.068$\pm$0.104 \\ 
        \textit{i3.large} & 0.181$\pm$0.496 \\ 
        \textit{m3.large} & 0.078$\pm$0.088 \\ 
        \textit{m3.medium} & 0.063$\pm$0.051 \\ 
        \textit{m4.large} & 0.052$\pm$0.054 \\ 
        \textit{r3.large} & 0.081$\pm$0.082 \\ 
        \textit{r4.large} & 0.080$\pm$0.055 \\ 
        \hline
        \end{tabular}
        \caption{Mean and Standard Deviation results of all instance types in the EU over 60 days (to 3dp)}
        \label{table:1}
    \begin{tabular}{ |c|c|c|c| } 
        \hline
        Instance Type & Mean$\pm$Standard Deviation(\$p/h)\\
        \hline
        \textit{c3.large} & 0.074$\pm$0.104 \\
        \textit{c4.large} & 0.079$\pm$0.107 \\ 
        \textit{i3.large }& 0.081$\pm$0.115 \\ 
        \textit{m3.large }& 0.088$\pm$0.102 \\ 
        \textit{m3.medium} & 0.063$\pm$0.055 \\ 
        \textit{m4.large }& 0.084$\pm$0.118 \\ 
        \textit{r3.large }& 0.081$\pm$0.066 \\ 
        \textit{r4.large }& 0.090$\pm$0.127 \\ 
        \hline
        \end{tabular}
        \caption{Mean and Standard Deviation results of all instance types in the US over 60 days (to 3dp)}
        \label{table:2}
    \begin{tabular}{ |c|c|c|c| } 
        \hline
        Instance Type & Mean$\pm$Standard Deviation(\$p/h)\\
        \hline
        \textit{c3.large} & 0.076$\pm$0.050 \\
        \textit{c4.large} & 0.096$\pm$0.139 \\ 
        \textit{i3.large }& 0.087$\pm$0.167 \\ 
        \textit{m3.large }& 0.085$\pm$0.056 \\ 
        \textit{m3.medium} & 0.063$\pm$0.047 \\ 
        \textit{m4.large }& 0.089$\pm$0.067 \\ 
        \textit{r3.large }& 0.088$\pm$0.058 \\ 
        \textit{r4.large }& 0.076$\pm$0.049 \\ 
        \hline
        \end{tabular}
        \caption{Mean and Standard Deviation results of all instance types in AP over 60 days (to 3dp)}
        \label{table:3}
    \begin{tabular}{ |c|c|c|c| } 
        \hline
        Instance Type & Mean$\pm$Standard Deviation(\$p/h)\\
        \hline
        \textit{c4.large} & 0.043$\pm$0.040 \\ 
        \textit{i3.large} & 0.072$\pm$0.045 \\ 
        \textit{m4.large} & 0.027$\pm$0.032 \\ 
        \textit{r4.large} & 0.029$\pm$0.033 \\ 
        \hline
        \end{tabular}
        \caption{Mean and Standard Deviation results of all instance types in CA over 60 days (to 3dp)}
        \label{table:4}
\end{table}

There are 68 different instance types that can be launched on EC2. If we had obtained the data for all 68, this would have led to an extremely large dataset. We therefore decided to restrict the dataset based on the most popular instance types available: small, medium and large instance types \cite{channelfutureswebsite}. Small instance types are not possible to deploy as a Spot Instance request so our final dataset consisted of medium and large instances. We then randomly picked a variety of instances (of varying compute power) from each of the AWS EC2 instance type categories - General Purpose, Compute Optimised, Memory Optimised and Storage Optimised \cite{amazonawsinstancetypes}.

Our final selection of data included \emph{m3.medium}, \emph{m4.large}, \emph{c4.large}, \emph{c3.large}, \emph{r3.large}, \emph{r4.large} and \emph{i3.large} instance types. The total pricing data over the 60 days for all the above instances was 4,112,000 data points.

\section{ANALYSES}

We focused on exploring the pricing data in relation to i) the \textit{instance type} and ii) the \textit{availability zone}.

We ran three main analyses.

\begin{itemize}
    \item First, we ran an average price analysis of all instance types within an AWS region over the full 60 days.
    \item Second, we ran an average price analysis for each instance in our list of instance types for every AWS region.
    \item Finally, we ran a histogram analysis plotting the frequency of all price points in a particular AZ for each of the instances deployed in that zone.
\end{itemize}

We broke down the first two analyses by day of the week and by hour in the day - showing the mean price across both these time metrics. We also calculated the standard deviation and related the average price analysis to these in order to examine the price volatility. In this context, volatility means the propensity for the price to change across a time metric. 

The lower the standard deviation, the lower the volatility. This meant that the data was closer to the mean and that we could be more confident in the mean price metrics reported. 

Conversely, the higher the standard deviation, the less confident we could be in the price's reliability.
 
 In the hour in the day analysis, we divided the day into 5 portions (00:00 to 05:00, 05:00-10:00, 10:00-15:00, 15:00-20:00 and 20:00-23:59). On the weekly analysis, we divided the week into 7 portions with 0 standing for Monday and 6 for Sunday.

\section{RESULTS}

\begin{figure*}[!htb]
\begin{center}
\subfigure[Average Pricing Per Hour: EU]{\includegraphics[width=0.35\textwidth]{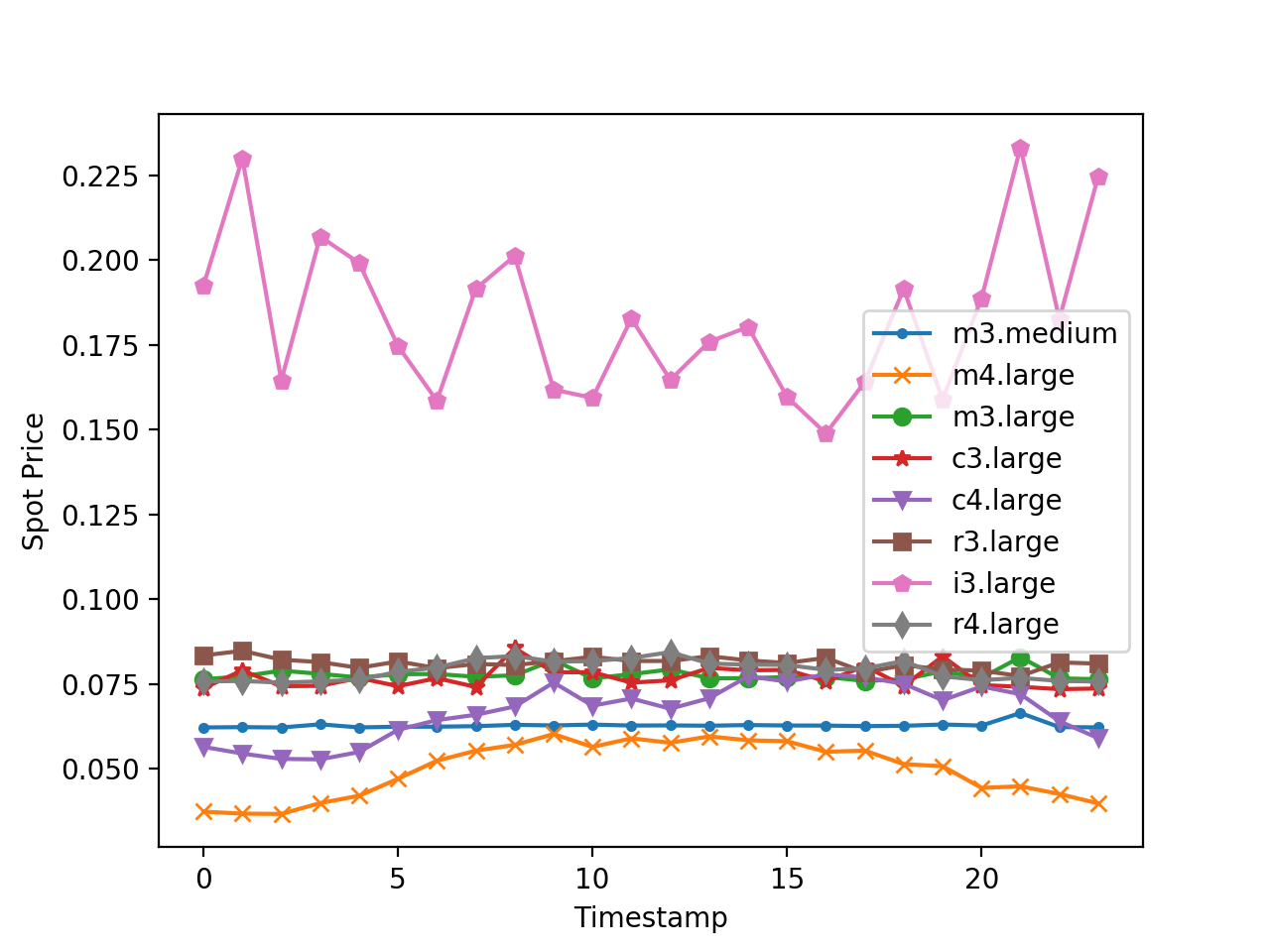} \label{fig:eu-all-instances-per-hour}}
\subfigure[Average Pricing Per Hour: US]{\includegraphics[width=0.35\textwidth]{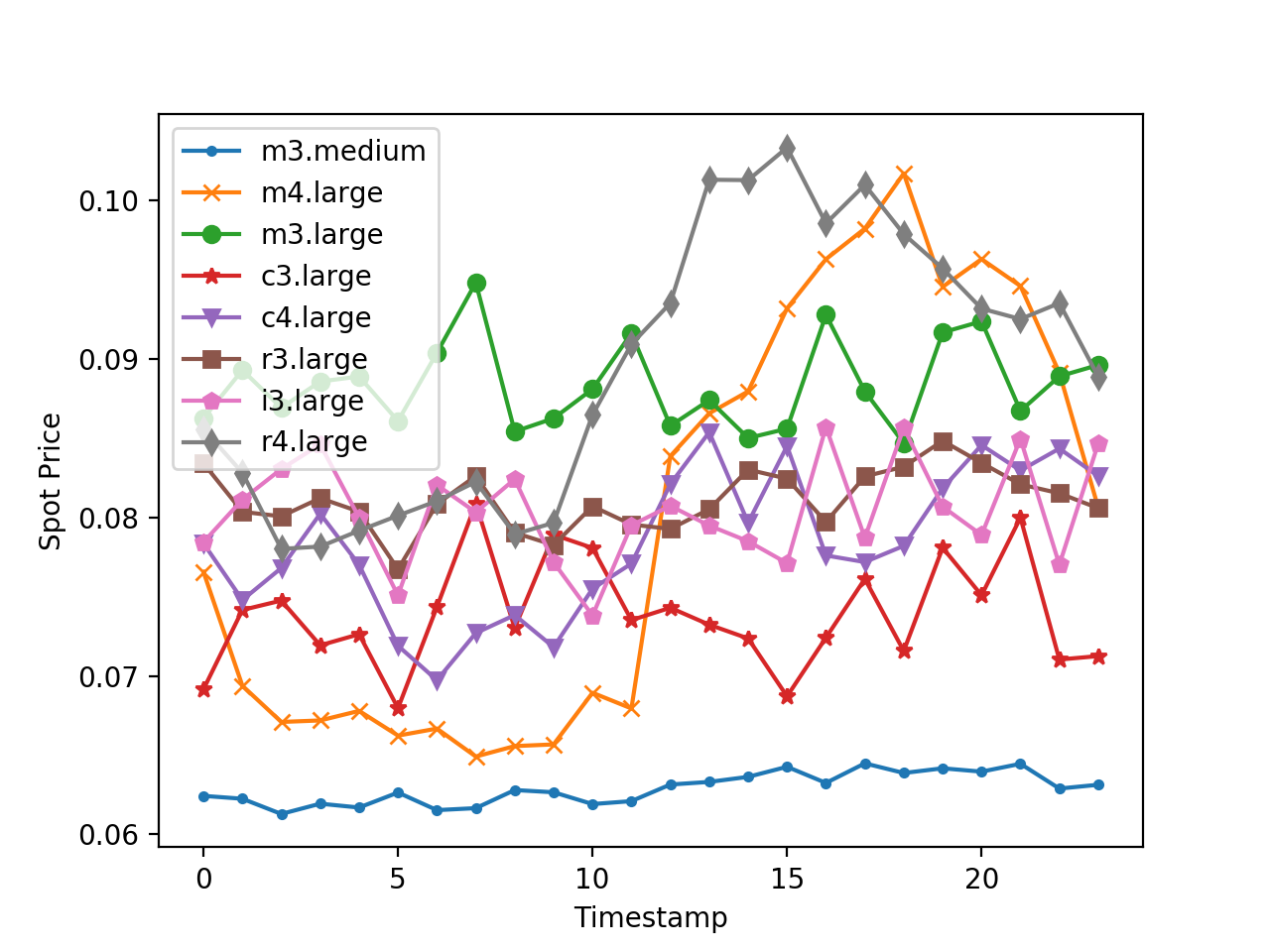} \label{fig:us-all-instances-per-hour}}
\subfigure[Average Pricing Per Hour: Asia-Pacific]{\includegraphics[width=0.35\textwidth]{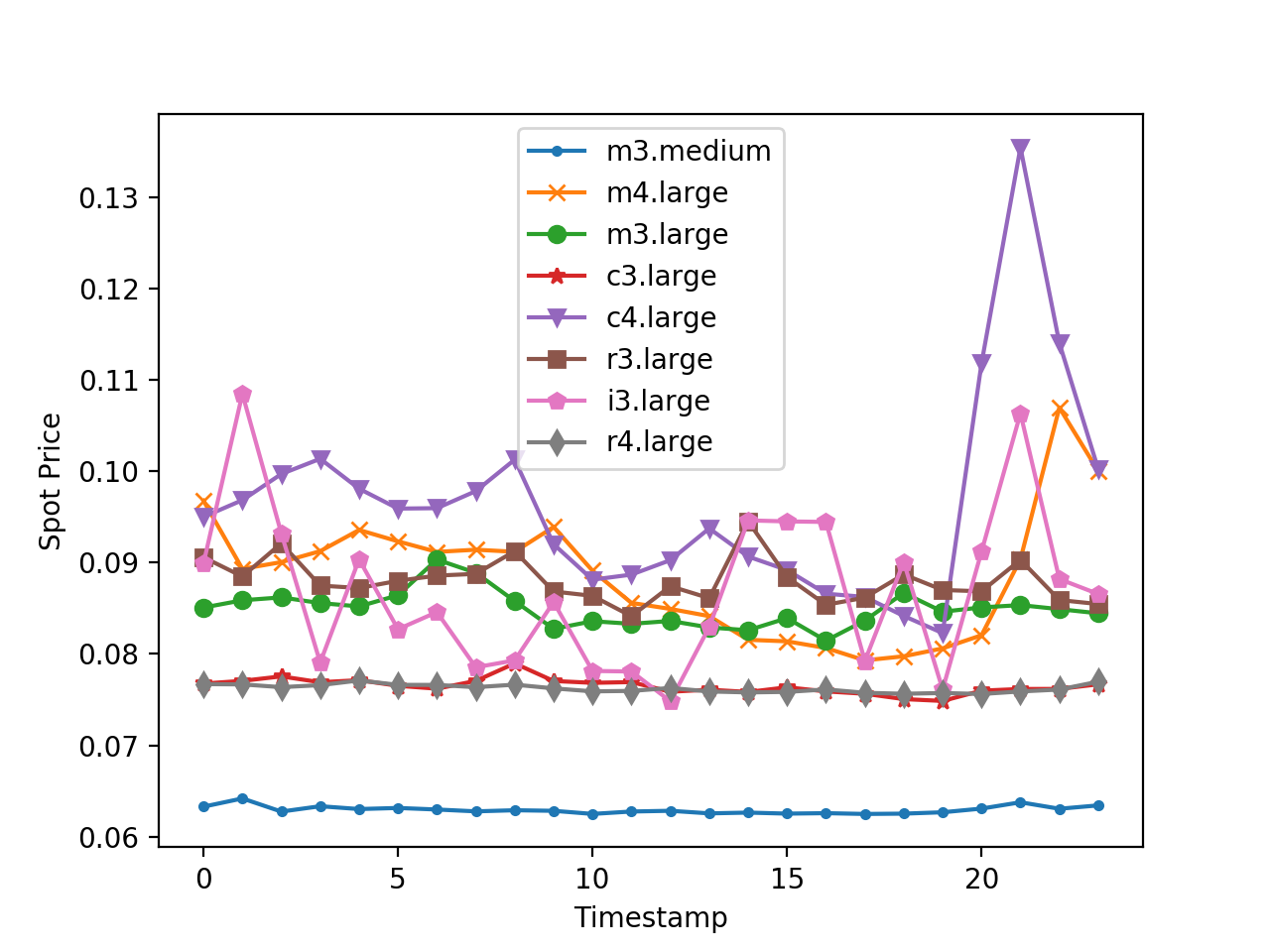} \label{fig:ap-all-instances-per-hour}}
\subfigure[Average Pricing Per Hour: Canada]{\includegraphics[width=0.35\textwidth]{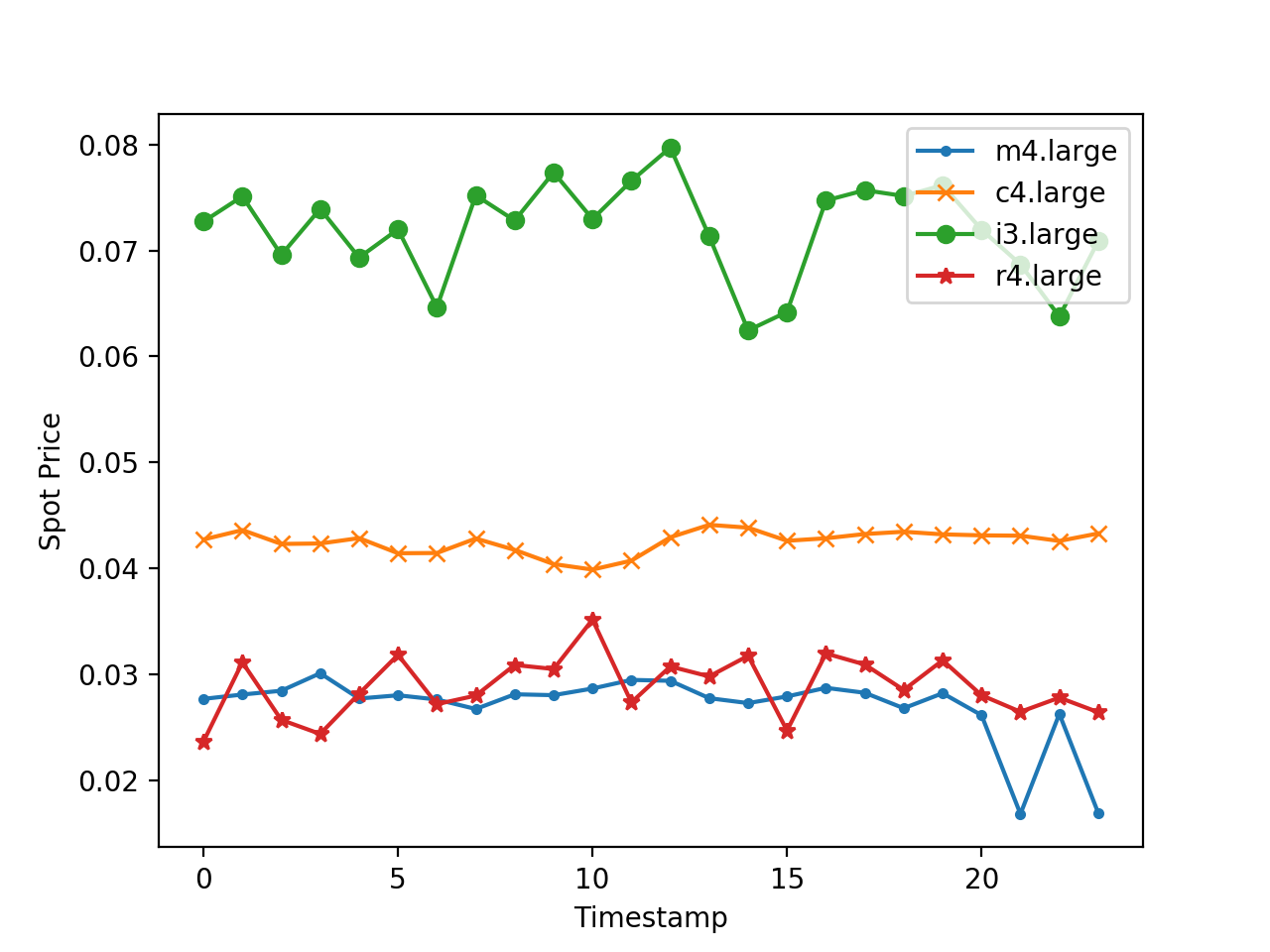} \label{fig:ca-all-instances-per-hour}}
\caption{Average Price Analysis of All Instances across all AWS regions: By Hour}
\label{fig:all-instances-per-hour-average-pricing}
\end{center}

\begin{center}
\subfigure[Average Pricing By Day Of Week: EU]{\includegraphics[width=0.35\textwidth]{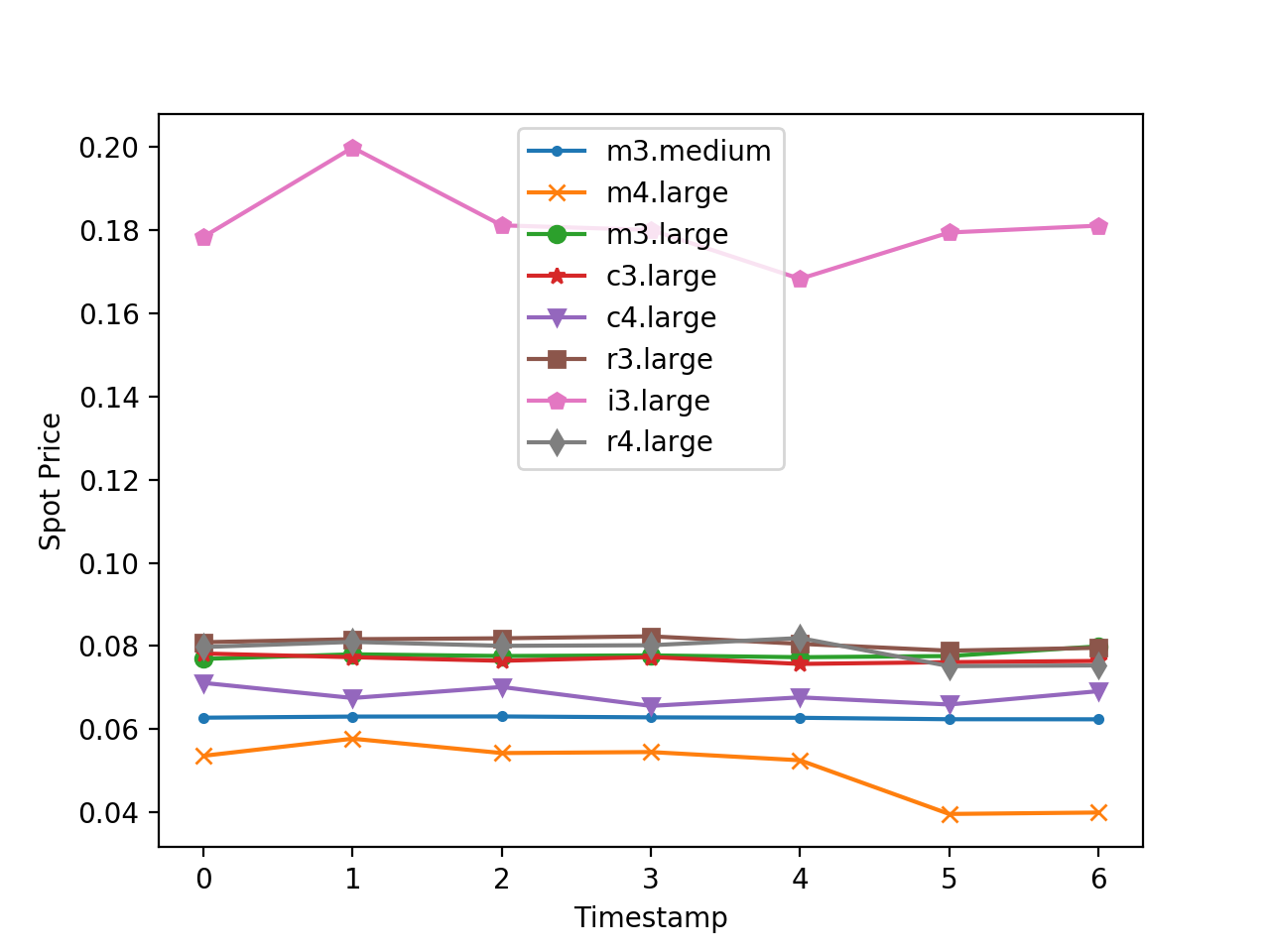} \label{fig:eu-all-instances-dow}}
\subfigure[Average Pricing By Day Of Week: US]{\includegraphics[width=0.35\textwidth]{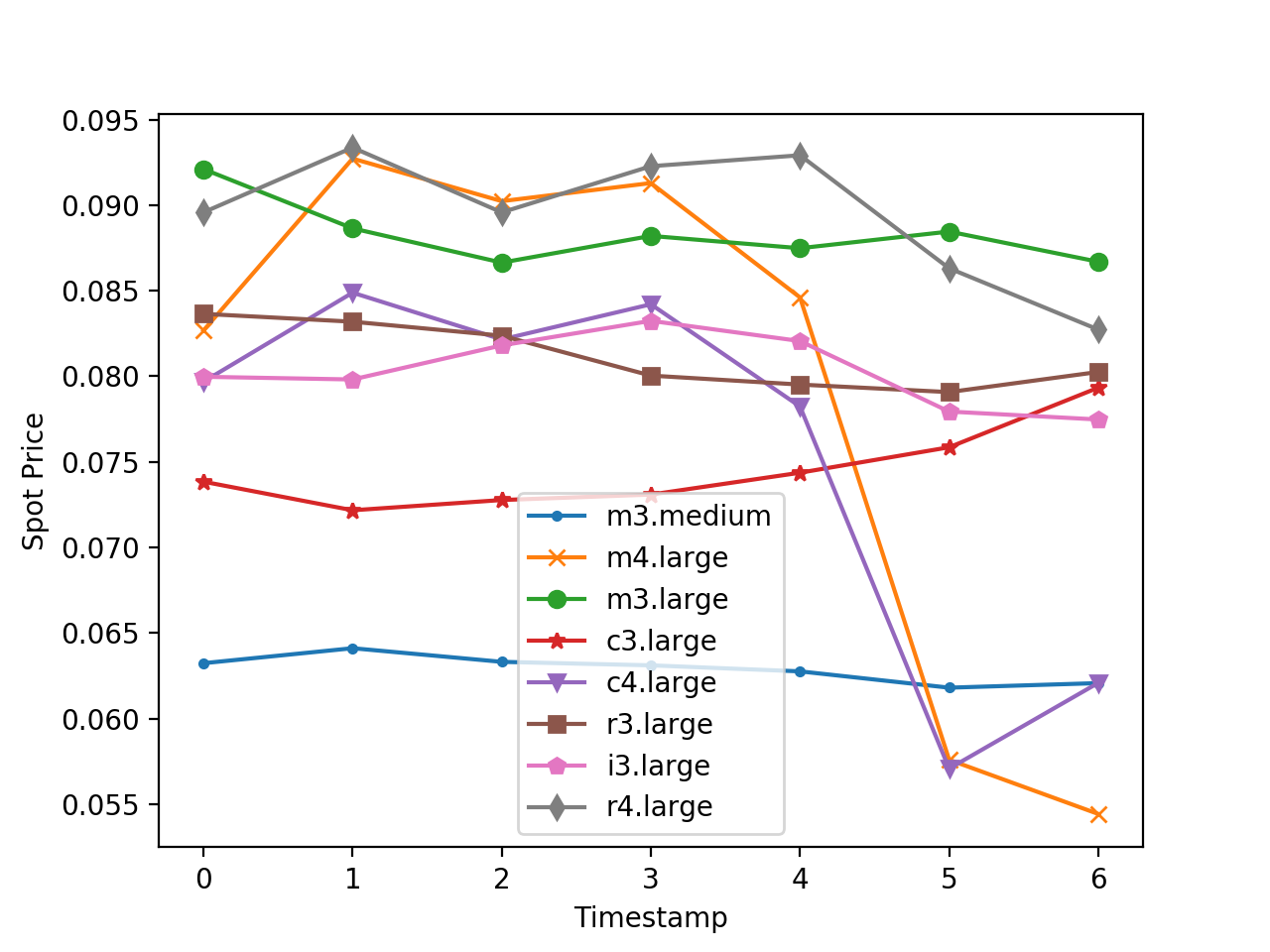} \label{fig:us-all-instances-dow}}
\subfigure[Average Pricing By Day Of Week: Asia-Pacific]{\includegraphics[width=0.35\textwidth]{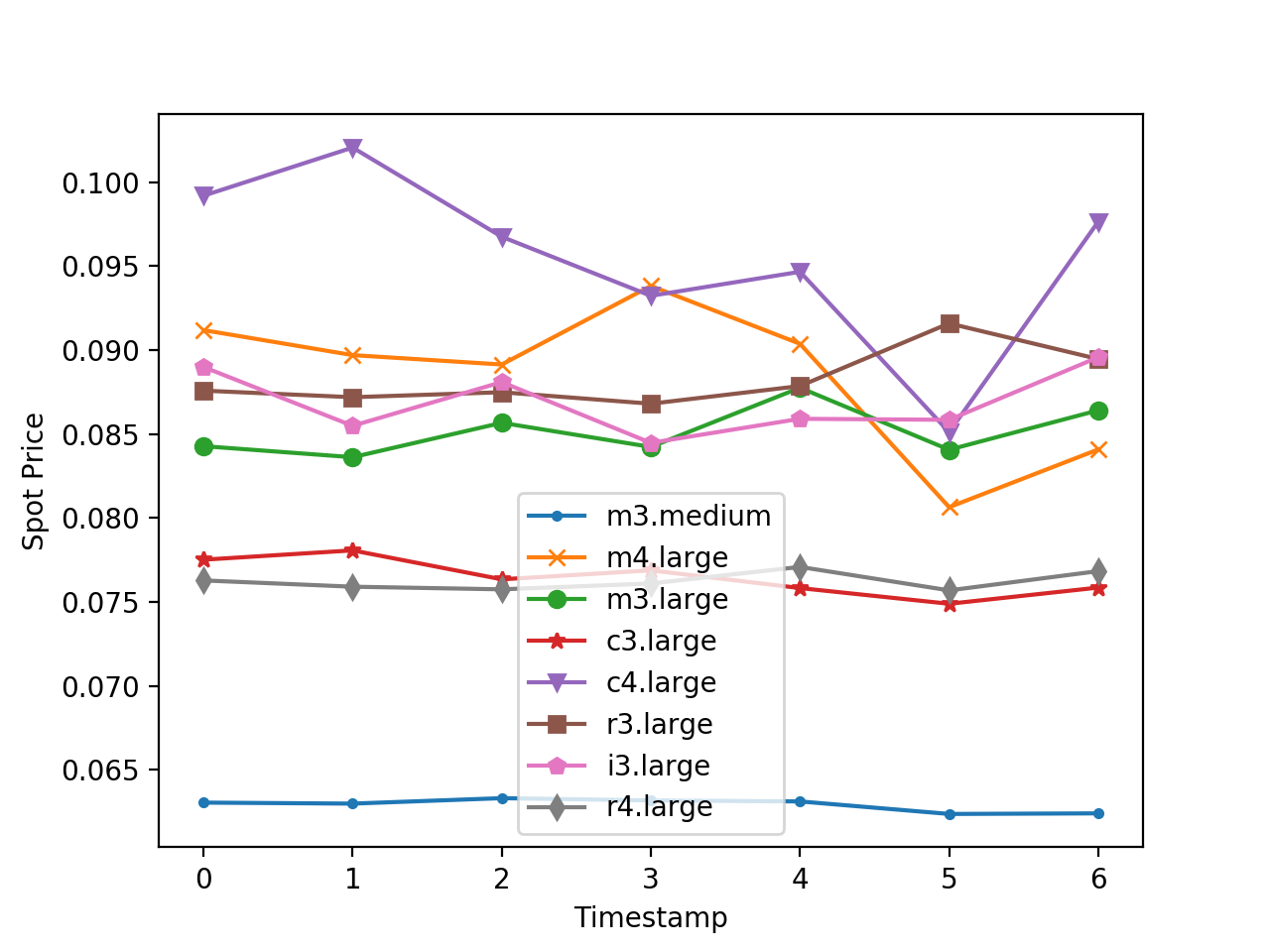} \label{fig:ap-all-instances-dow}}
\subfigure[Average Pricing By Day Of Week: Canada]{\includegraphics[width=0.35\textwidth]{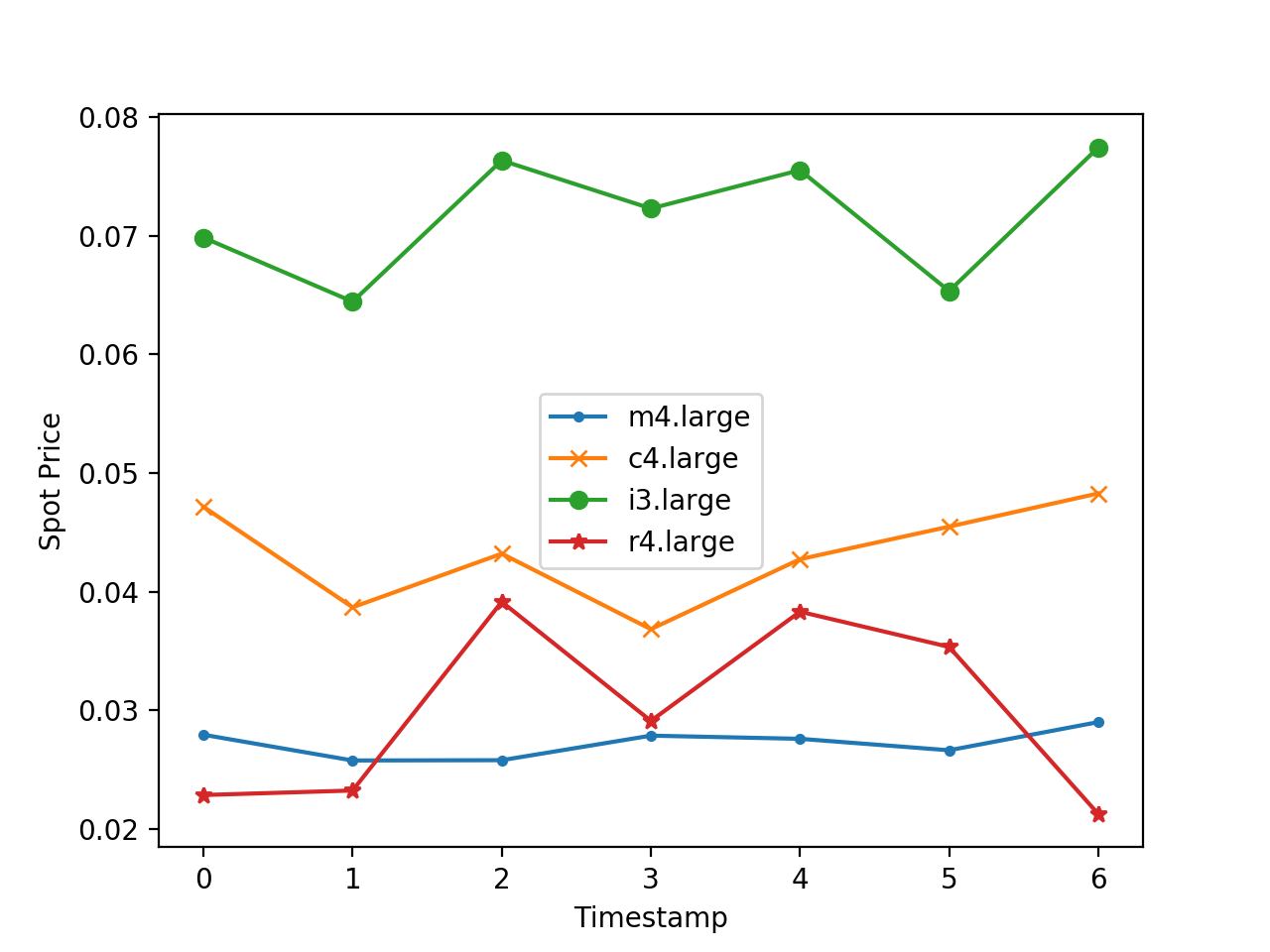} \label{fig:ca-all-instances-per-dow}}
\caption{Average Price Analysis of All Instances across all AWS regions: By Day of Week}
\label{fig:all-instances-per-dow-average-pricing}
\end{center}
\end{figure*}

\begin{figure*}[!htb]
\begin{center}

\subfigure[m3.medium]{\includegraphics[width=0.32\textwidth]{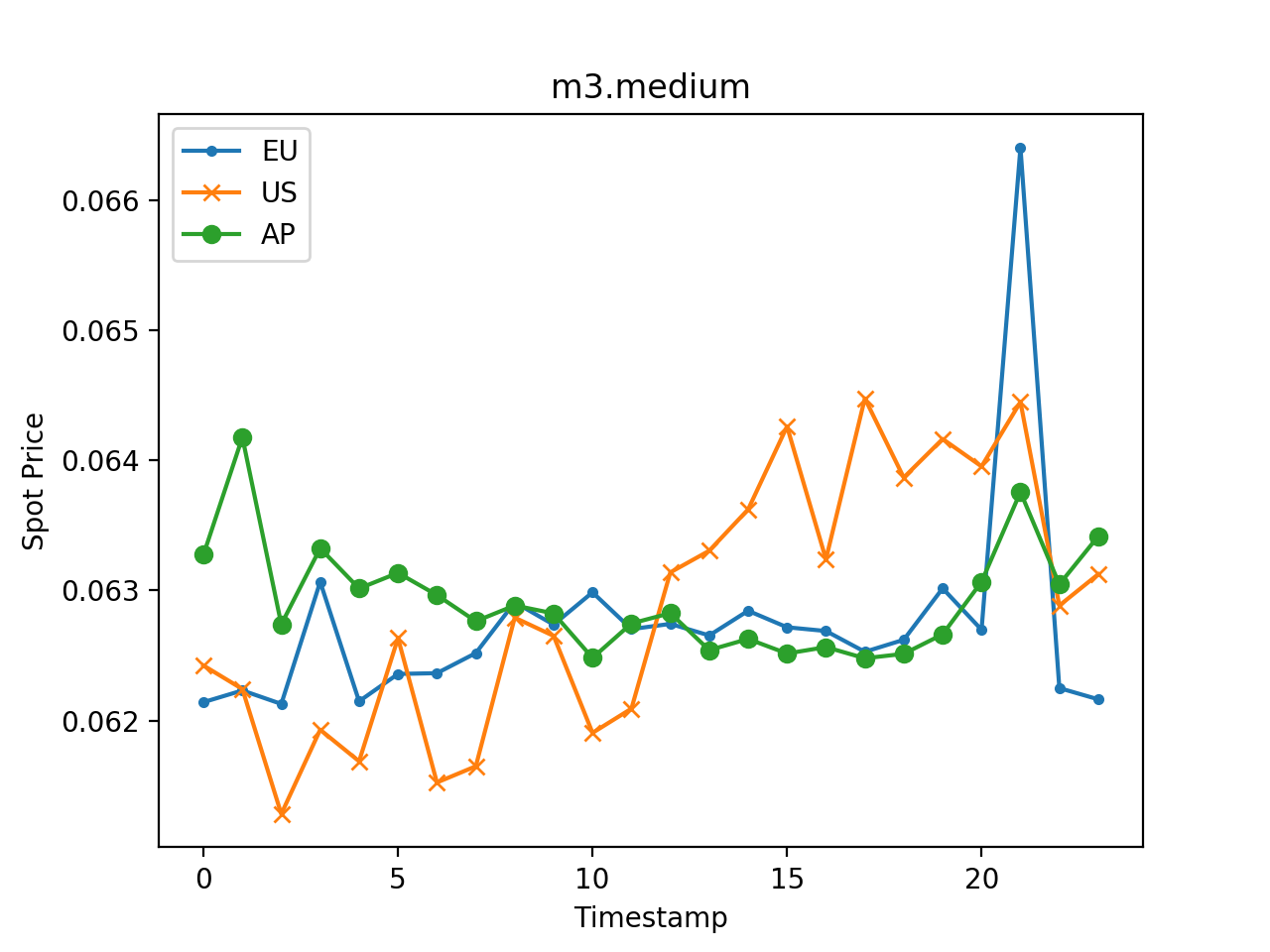} \label{fig:m3-medium-average-pricing-per-hour}}
\subfigure[m3.large]{\includegraphics[width=0.32\textwidth]{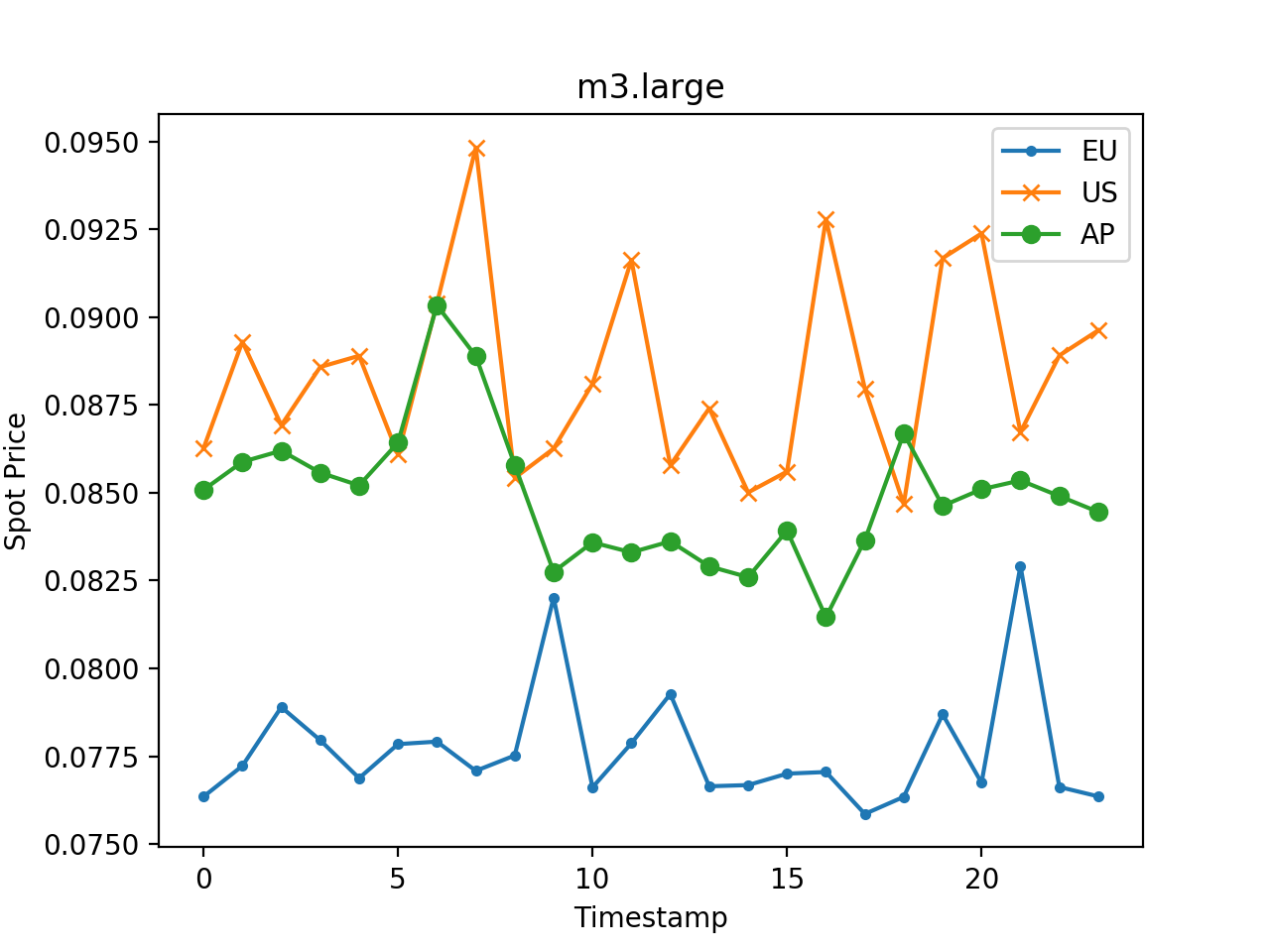} \label{fig:m3-large-average-pricing-per-hour}}
\subfigure[m4.large]{\includegraphics[width=0.32\textwidth]{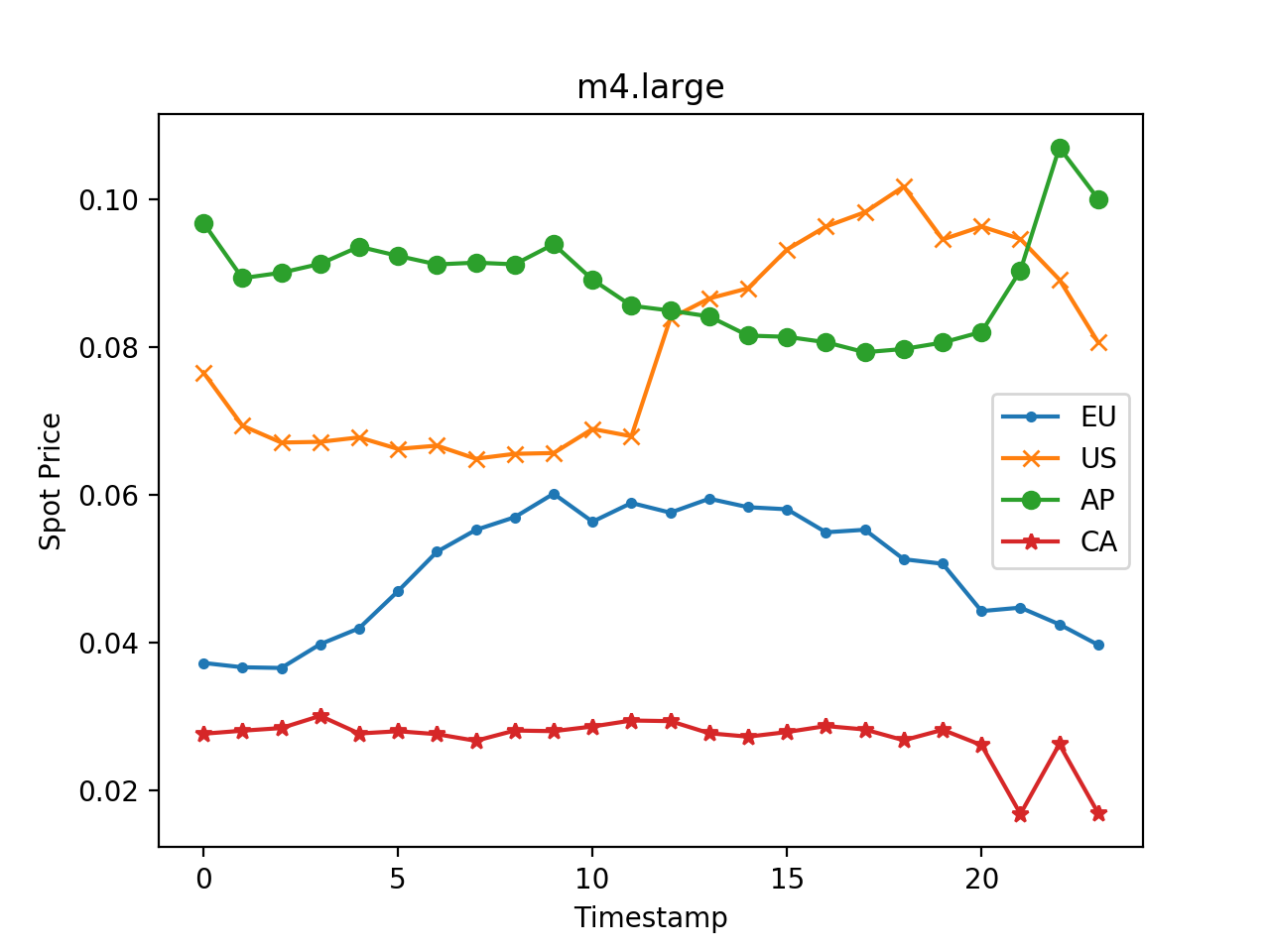} \label{fig:m4-large-average-pricing-per-hour}}

\subfigure[c4.large]{\includegraphics[width=0.32\textwidth]{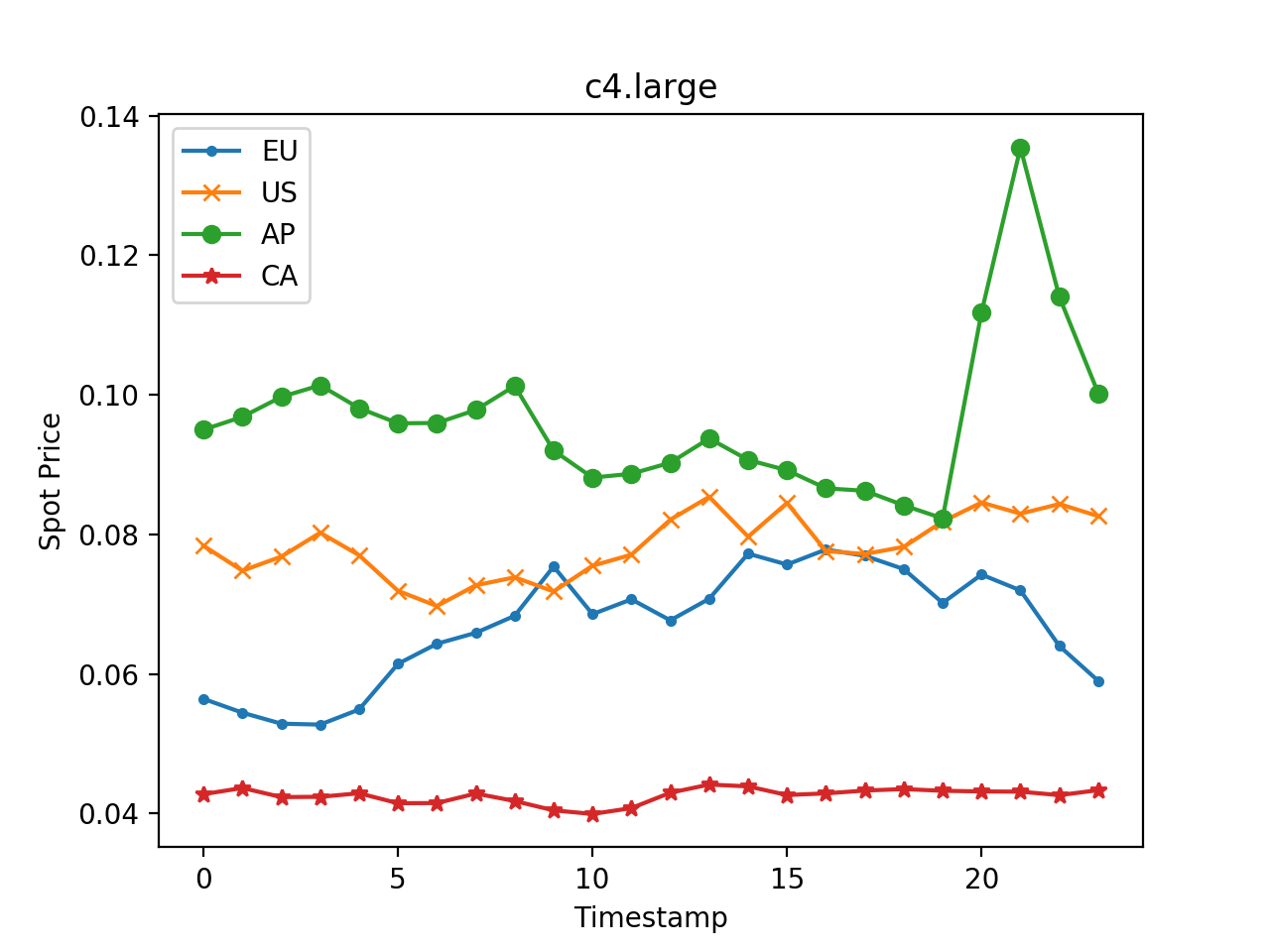} \label{fig:c4.large-average-pricing-per-hour}}
\subfigure[c3.large]{\includegraphics[width=0.32\textwidth]{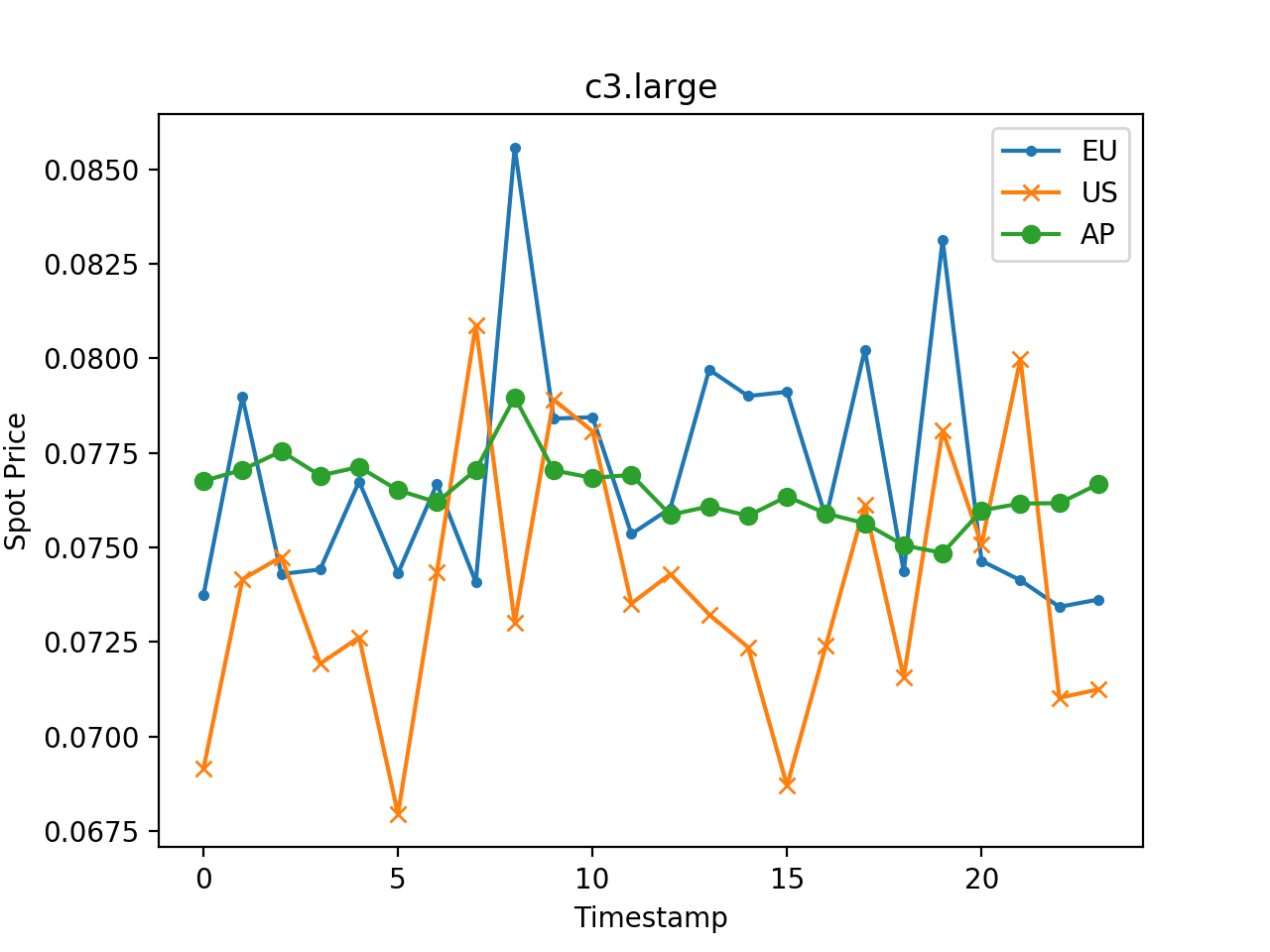} \label{fig:c3.large-average-pricing-per-hour}}
\subfigure[r3.large]{\includegraphics[width=0.32\textwidth]{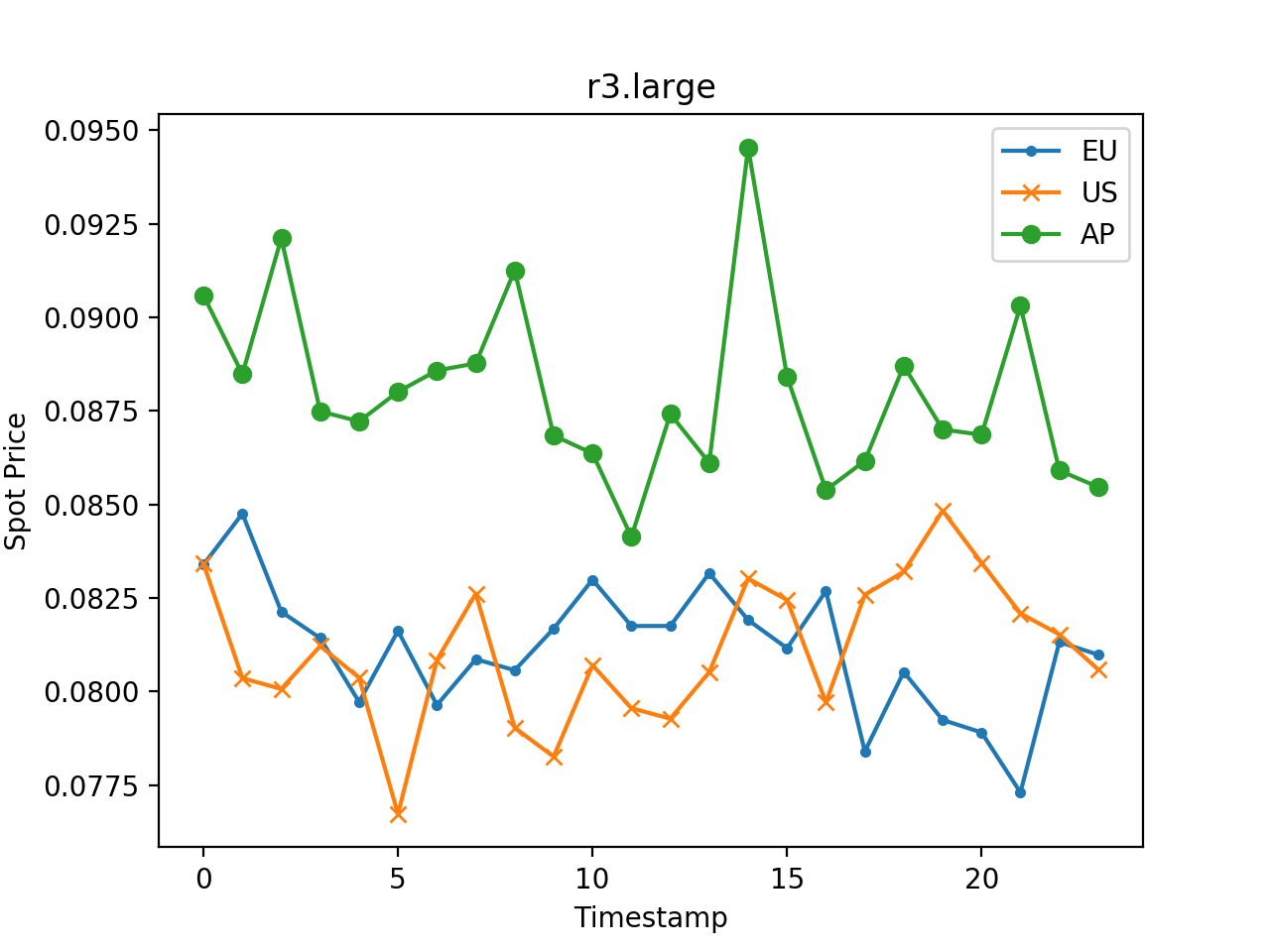} \label{fig:r3-large-average-pricing-per-hour}}

\subfigure[r4.large]{\includegraphics[width=0.32\textwidth]{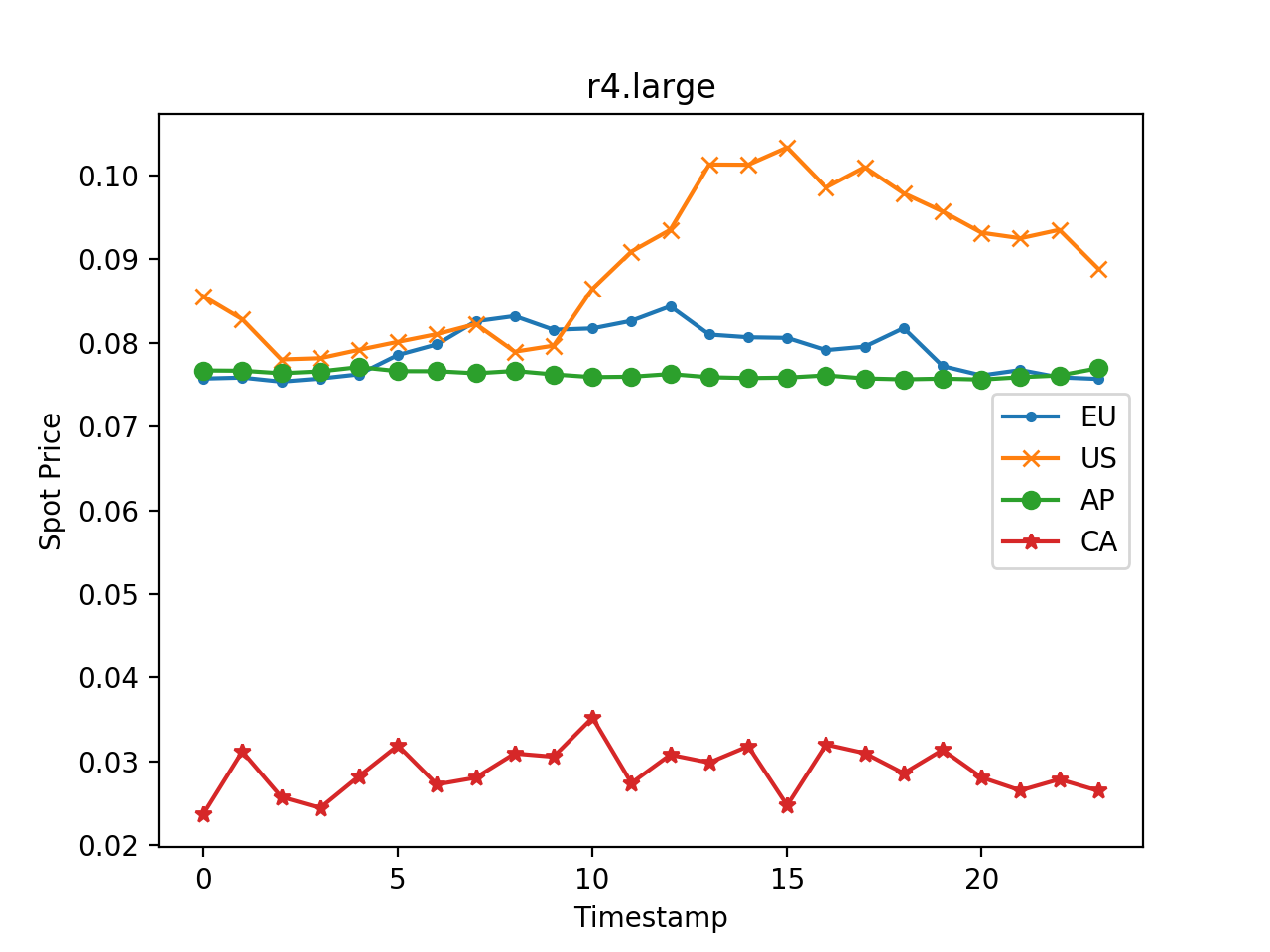} \label{fig:r4-large-average-pricing-per-hour}}
\subfigure[i3.large]{\includegraphics[width=0.32\textwidth]{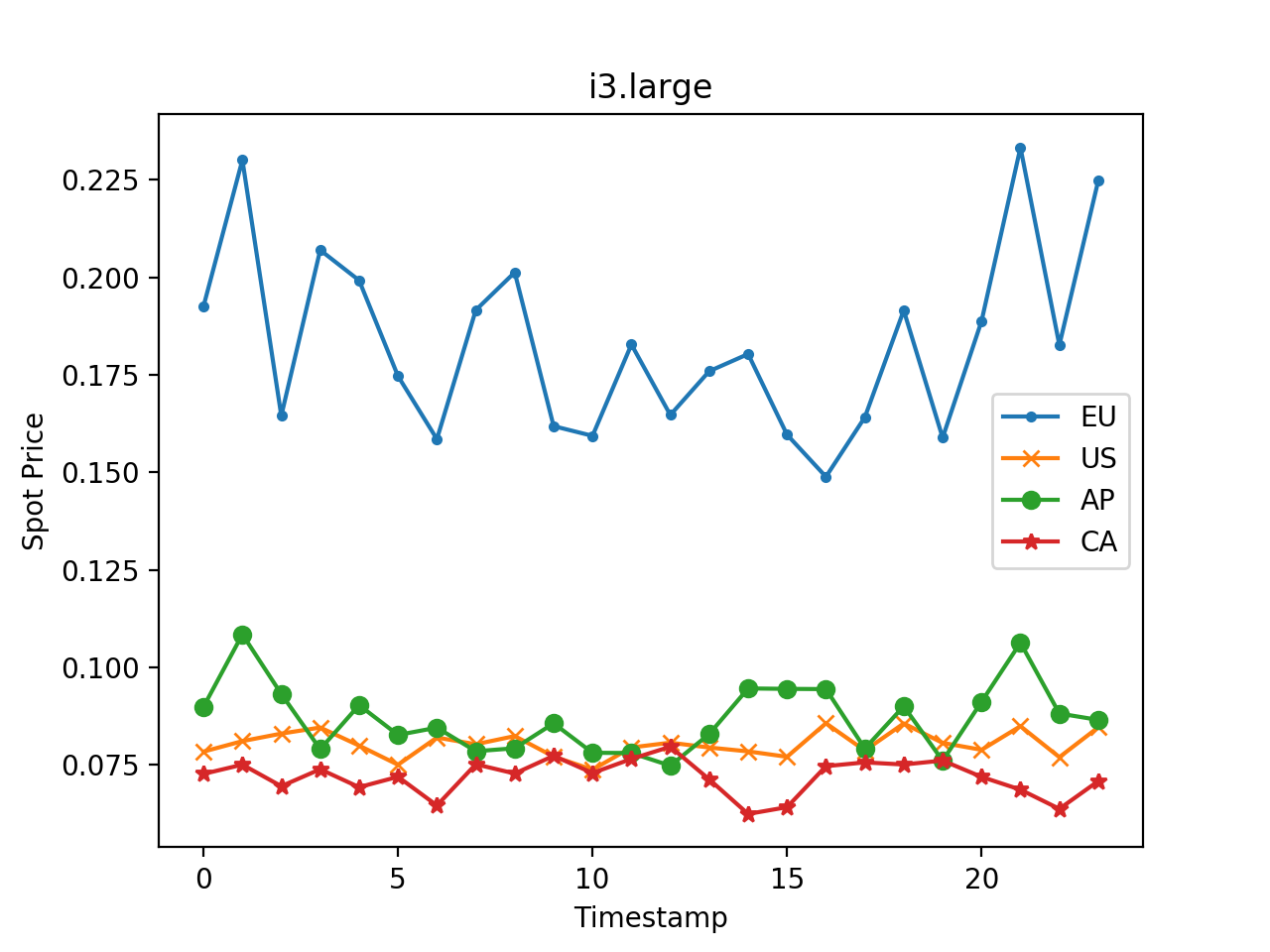} \label{fig:i3-large-average-pricing-per-hour}}

\caption{Average Price Analysis of Each Instance for every AWS region: By Hour}
\label{fig:each-instance-every-region-per-hour-average-pricing}
\end{center}
\end{figure*}

\begin{figure*}[!htb]
\begin{center}

\subfigure[m3.medium]{\includegraphics[width=0.32\textwidth]{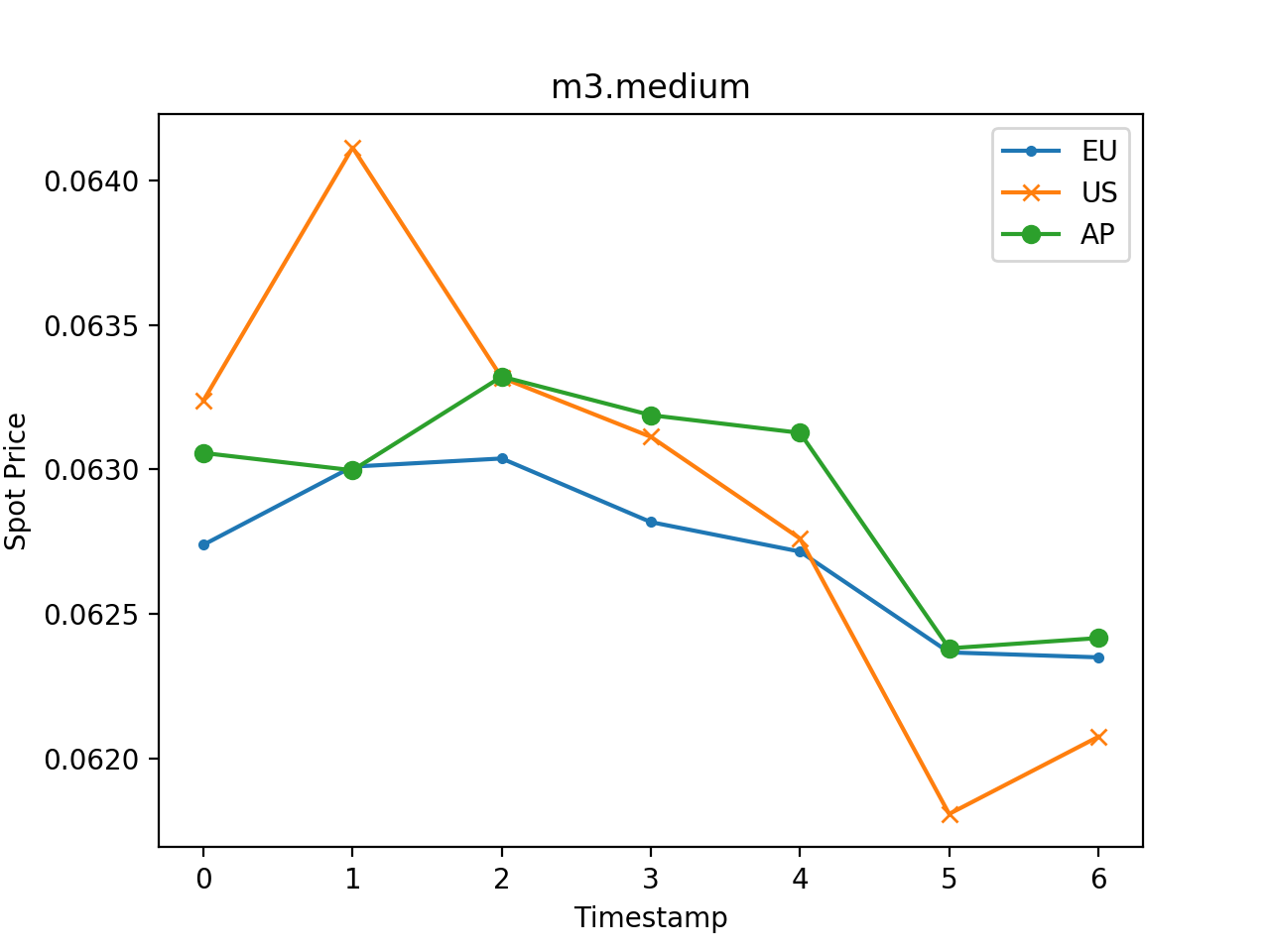} \label{fig:m3-medium-average-pricing-dow}}
\subfigure[m3.large]{\includegraphics[width=0.32\textwidth]{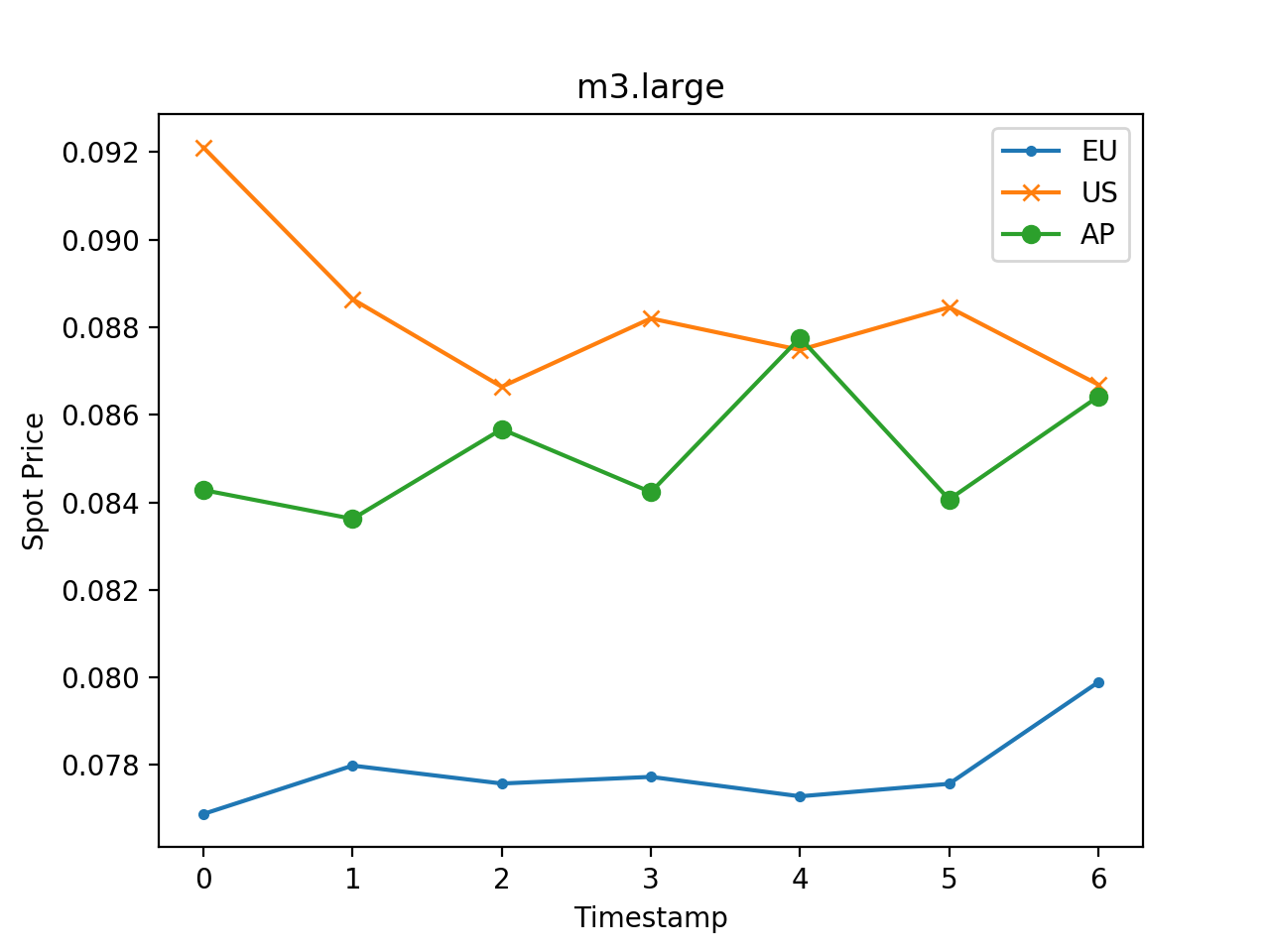} \label{fig:m3-large-average-pricing-dow}}
\subfigure[m4.large]{\includegraphics[width=0.32\textwidth]{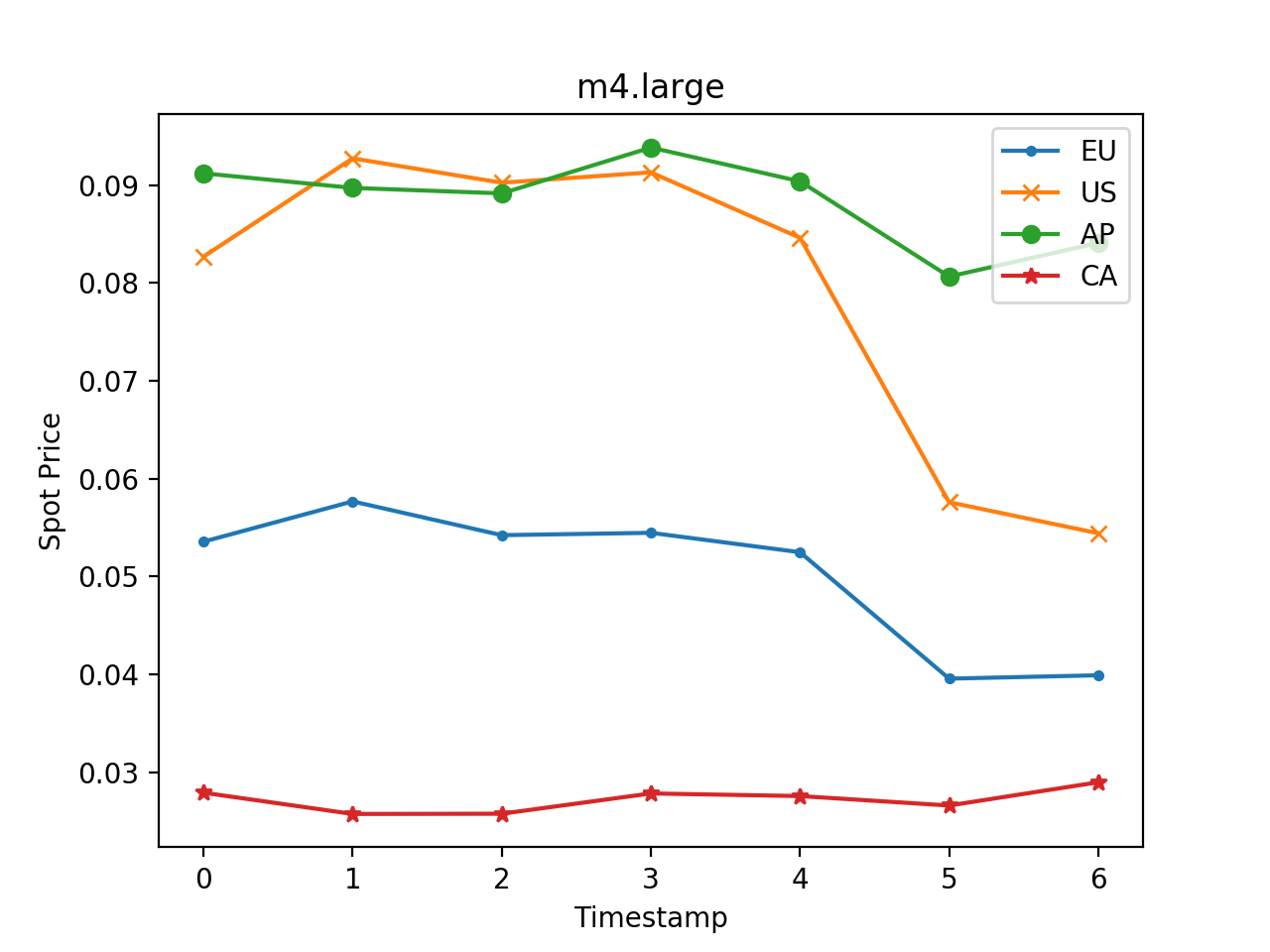} \label{fig:m4-large-average-pricing-dow}}

\subfigure[c4.large]{\includegraphics[width=0.32\textwidth]{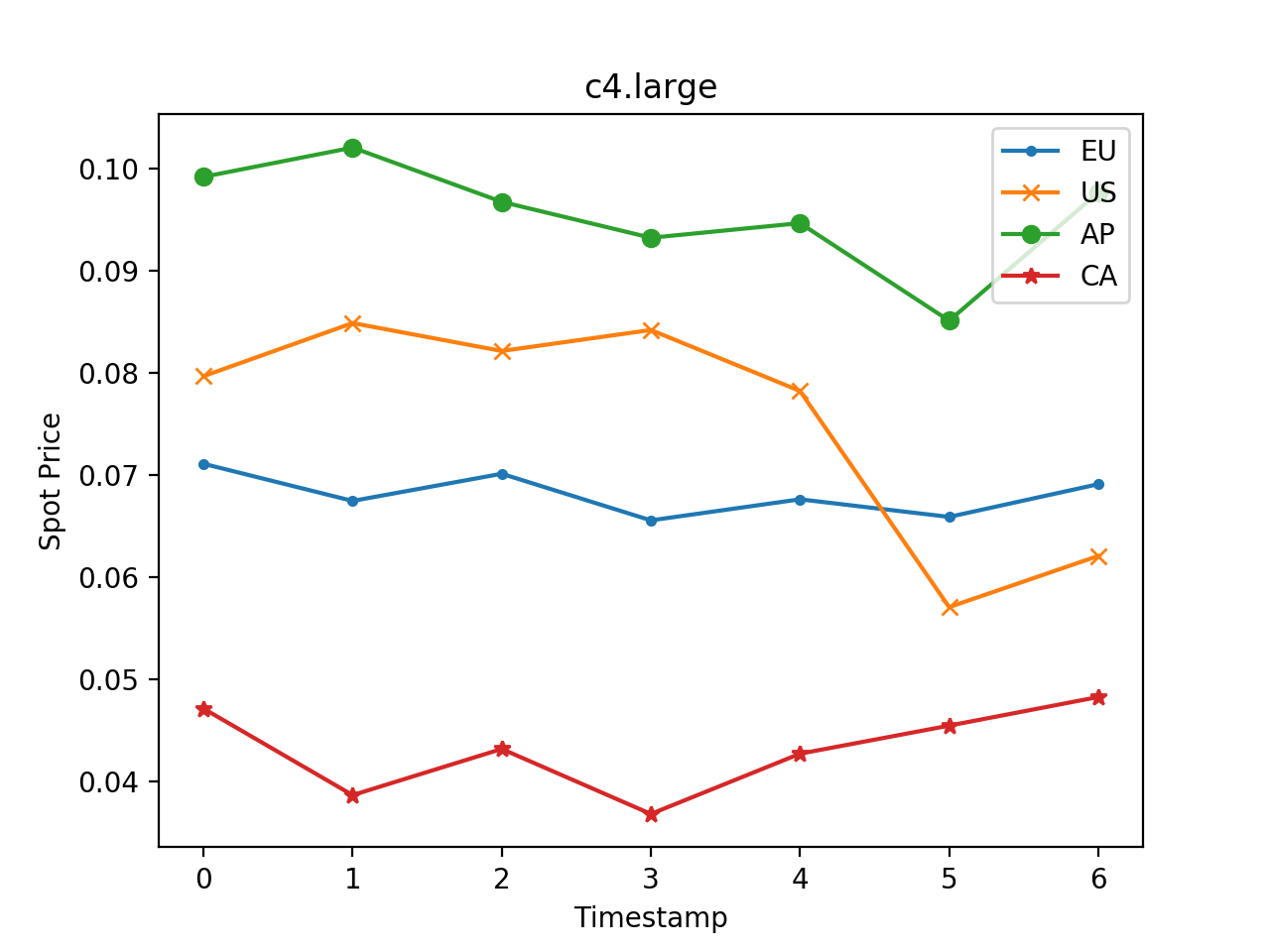} \label{fig:c4.large-average-pricing-dow}}
\subfigure[c3.large]{\includegraphics[width=0.32\textwidth]{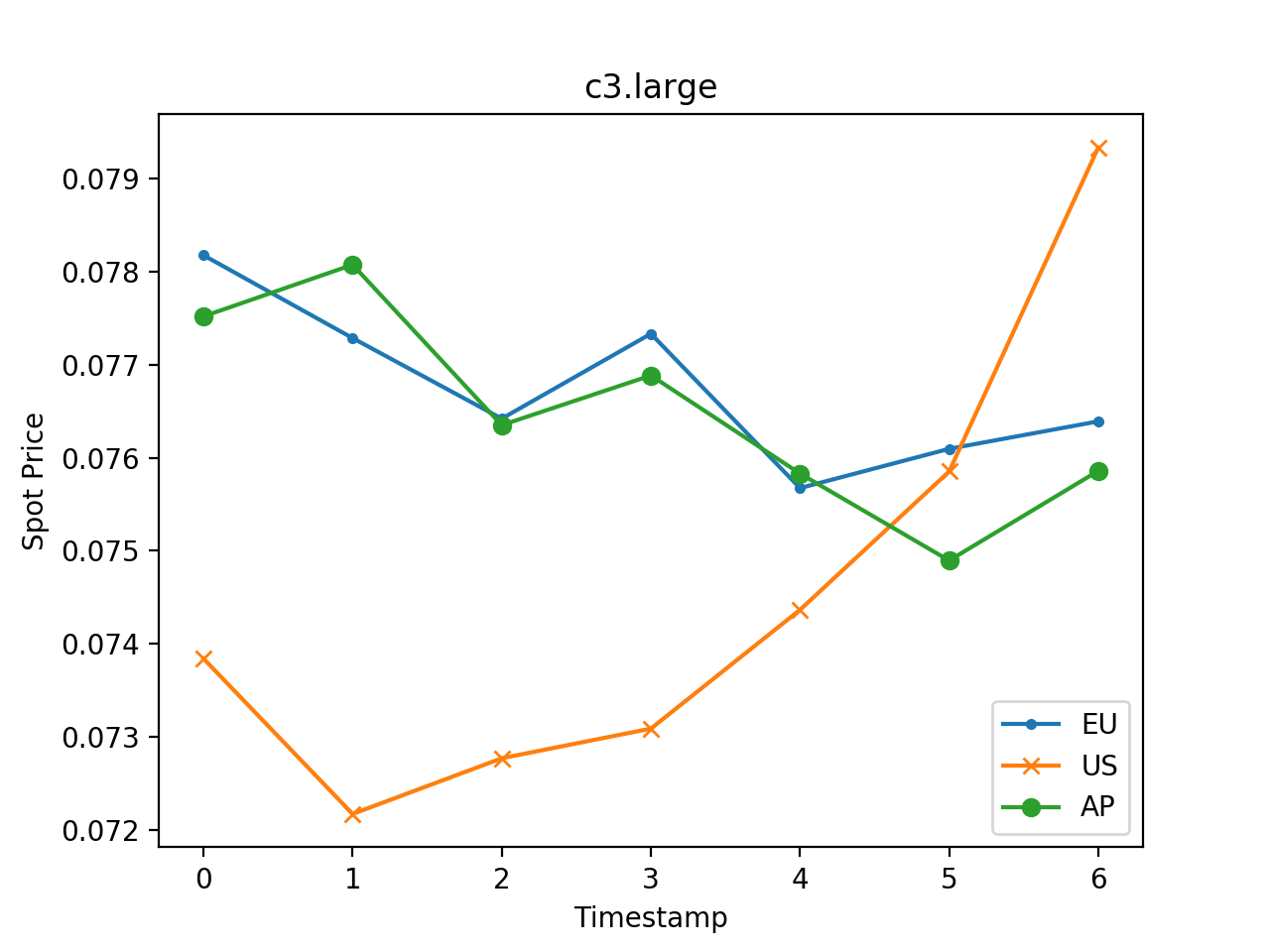} \label{fig:c3.large-average-pricing-dow}}
\subfigure[r3.large]{\includegraphics[width=0.32\textwidth]{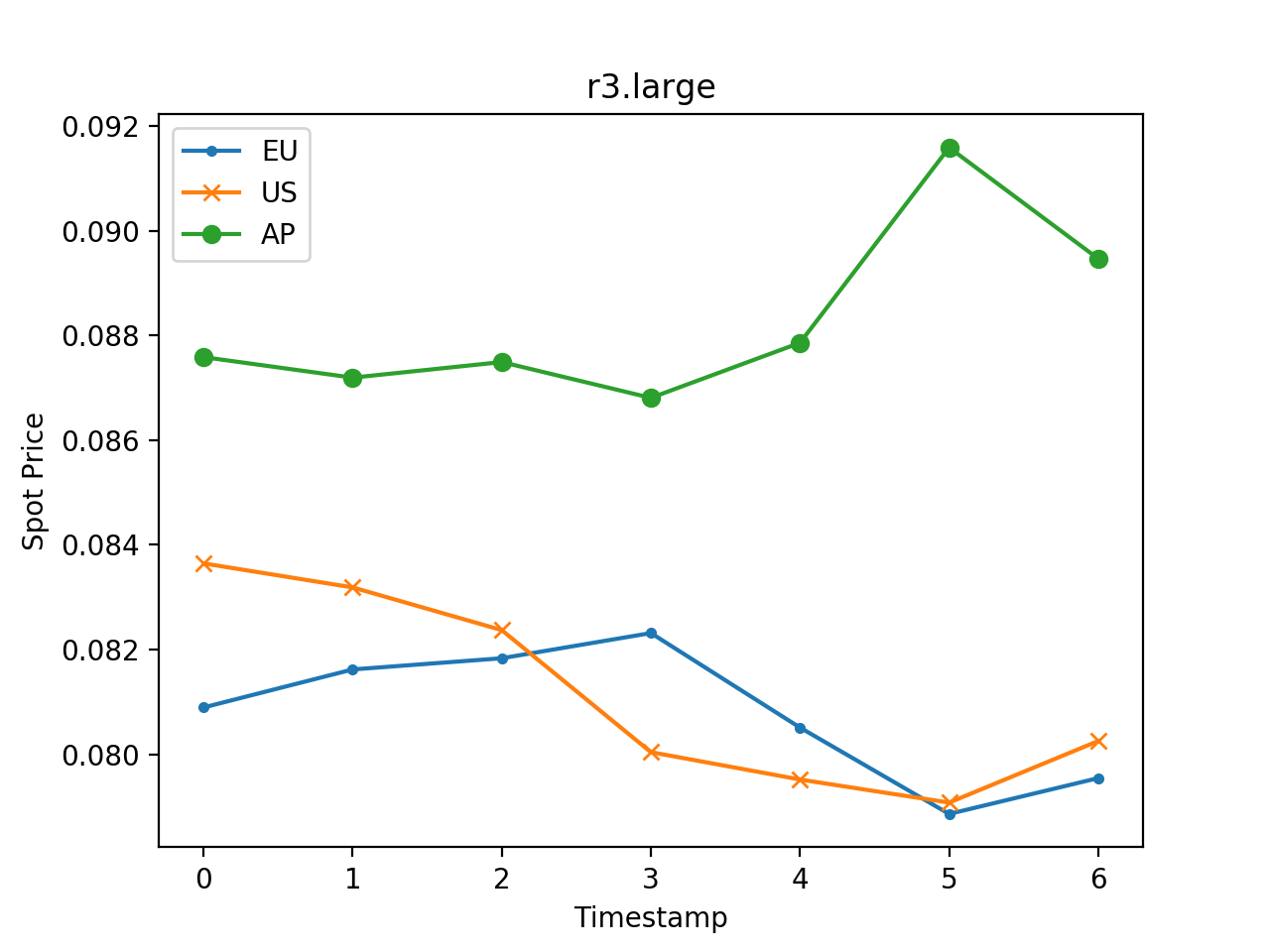} \label{fig:r3-large-average-pricing-dow}}

\subfigure[r4.large]{\includegraphics[width=0.32\textwidth]{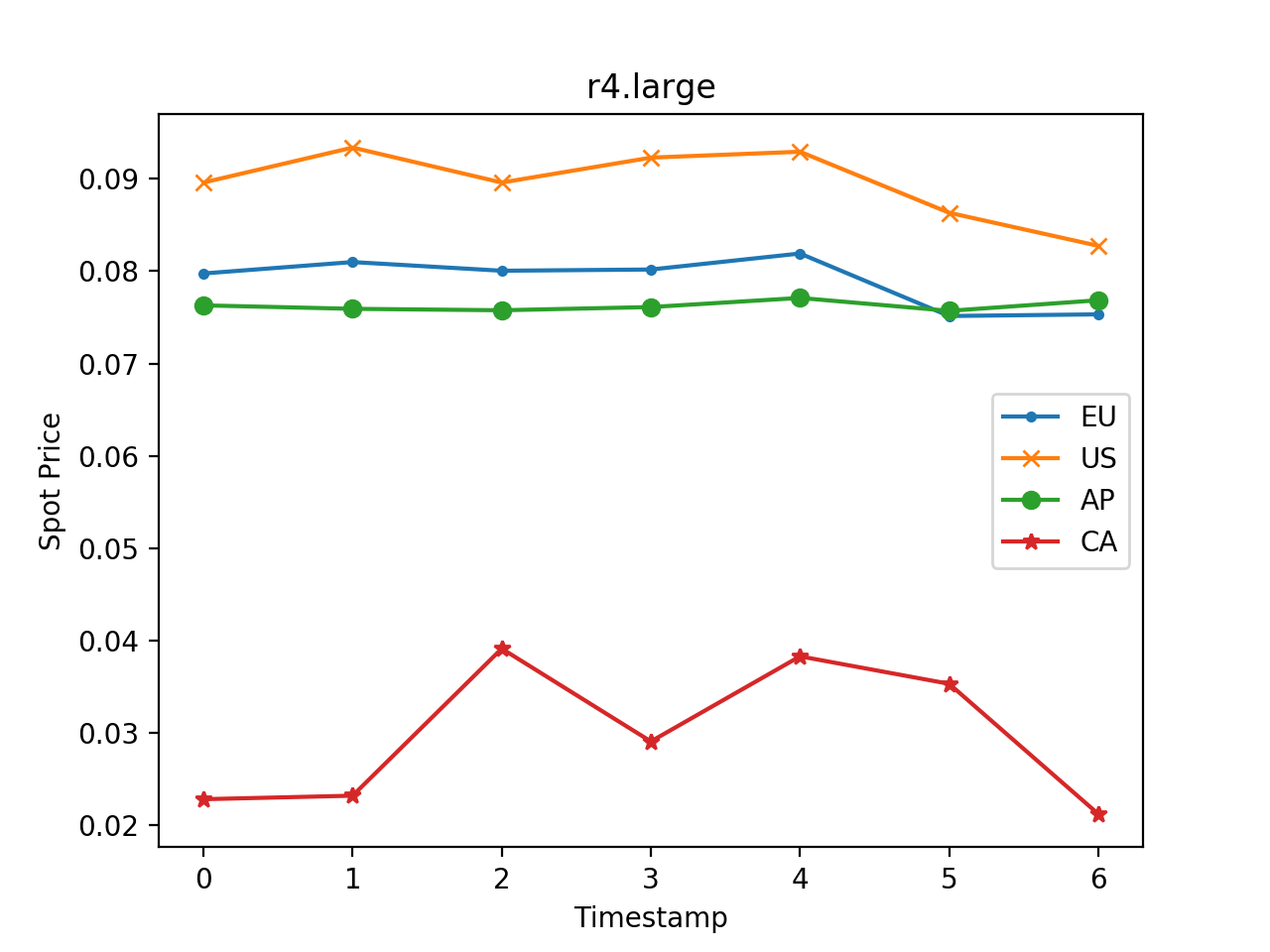} \label{fig:r4-large-average-pricing-dow}}
\subfigure[i3.large]{\includegraphics[width=0.32\textwidth]{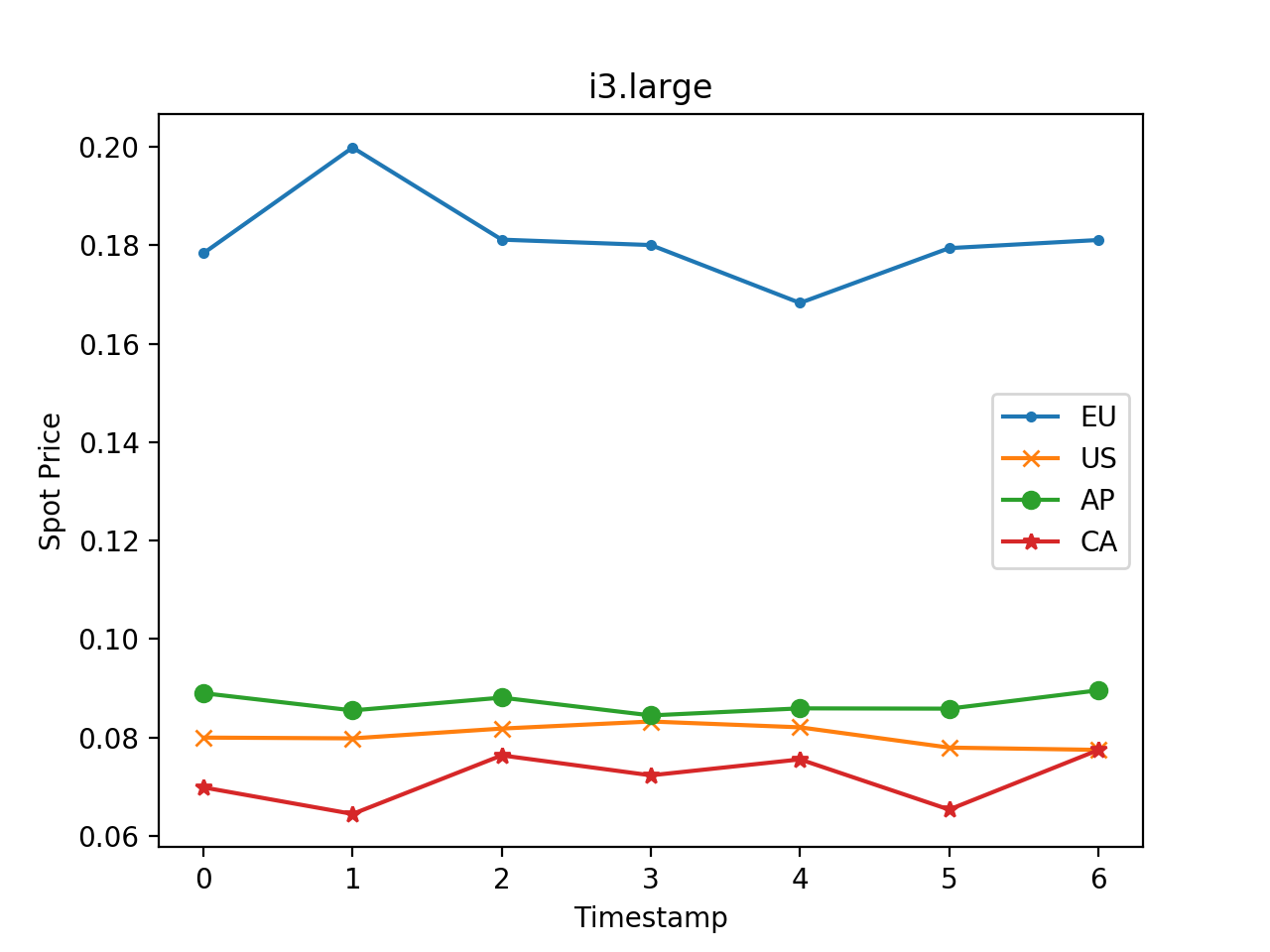} \label{fig:i3-large-average-pricing-dow}}
\caption{Average Price Analysis of Each Instance for every AWS region: By Day of Week}
\label{fig:each-instance-every-region-per-dow-average-pricing}
\end{center}
\end{figure*}

\begin{figure*}[!htp]
\begin{center}

\subfigure[eu-central-1a]{\includegraphics[width=0.32\textwidth]{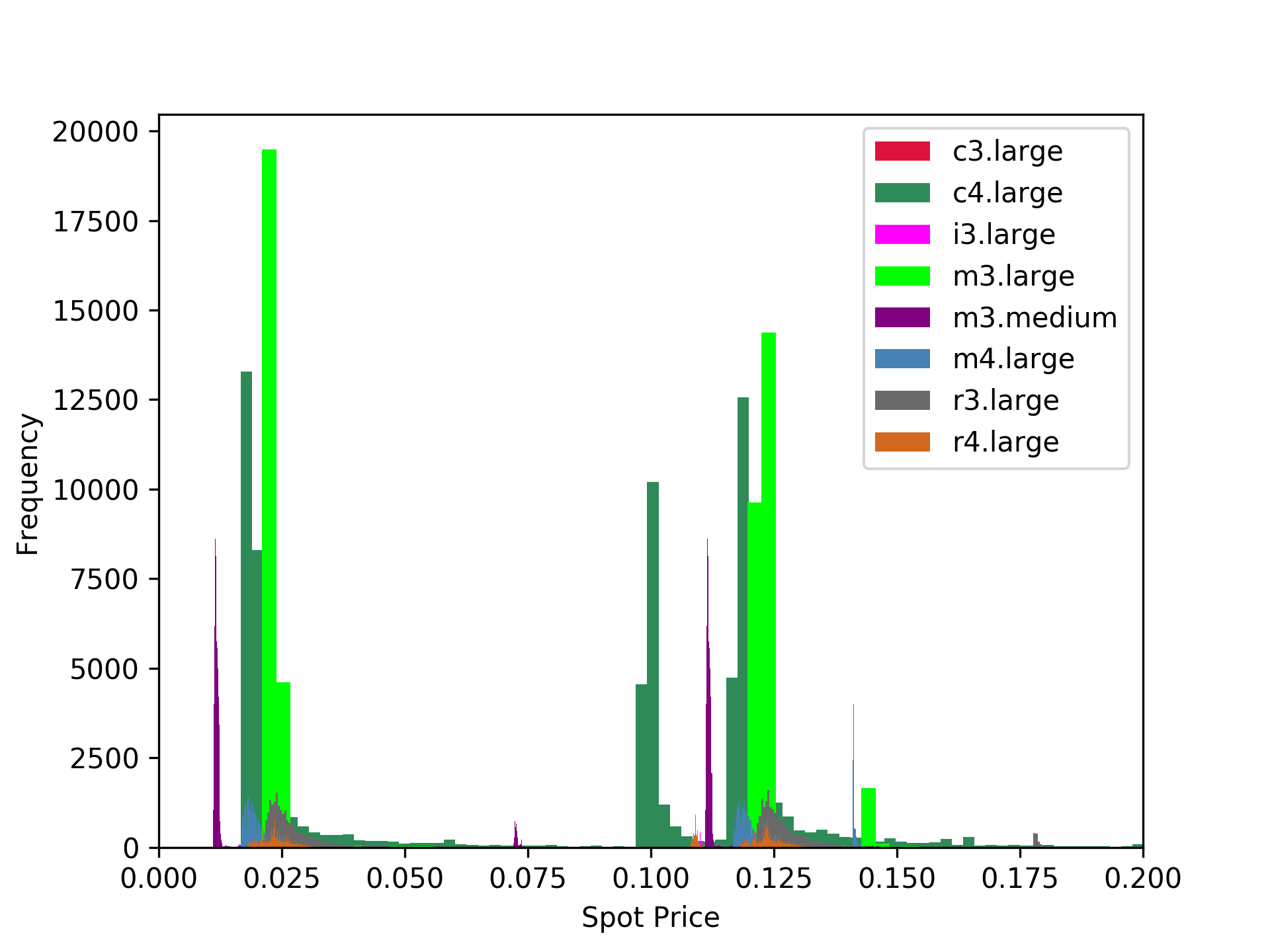} \label{fig:a1}}
\subfigure[eu-central-1b]{\includegraphics[width=0.32\textwidth]{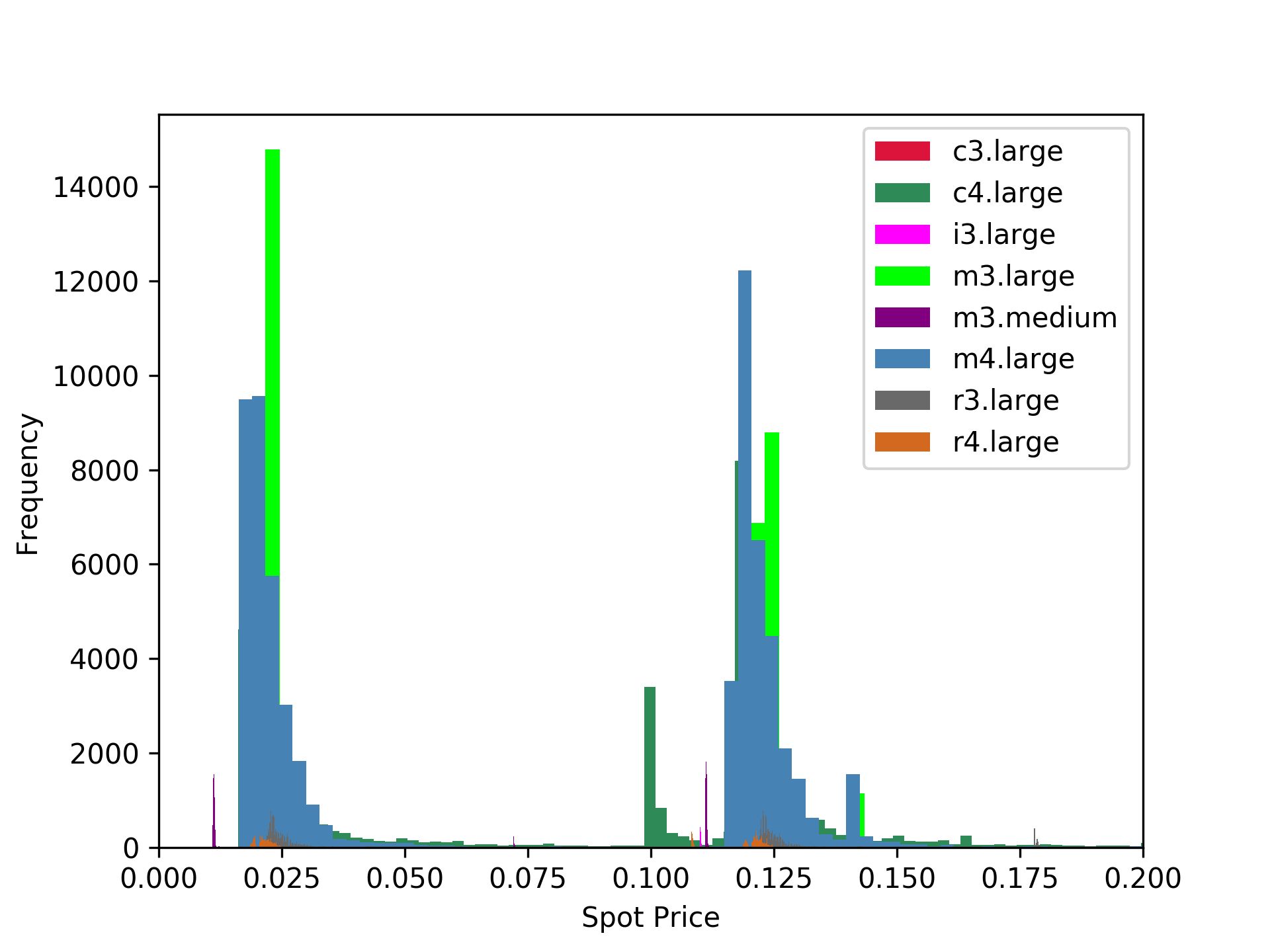} \label{fig:a2}}
\subfigure[eu-west-1a]{\includegraphics[width=0.32\textwidth]{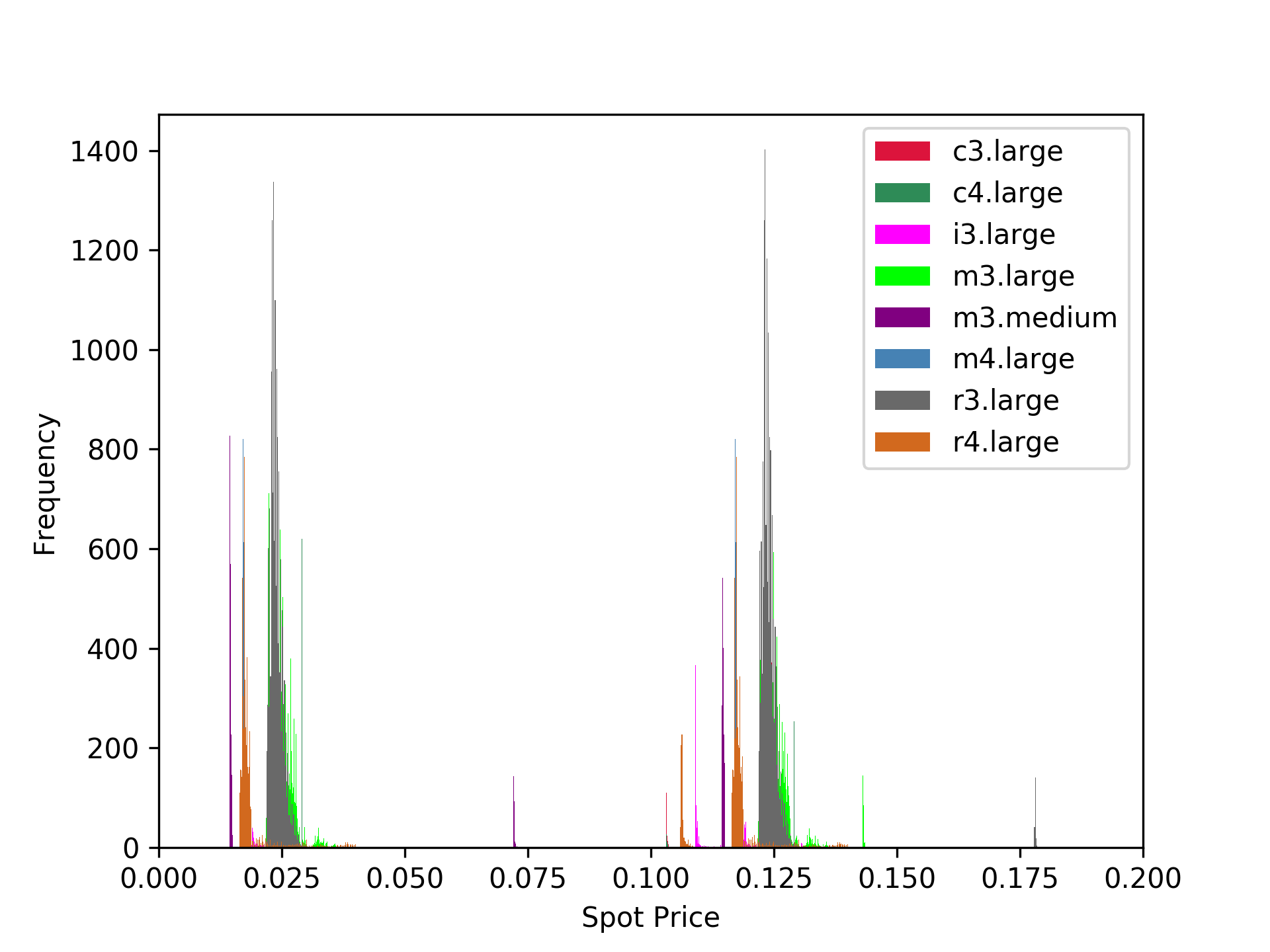} \label{fig:a3}}

\subfigure[eu-west-1b]{\includegraphics[width=0.32\textwidth]{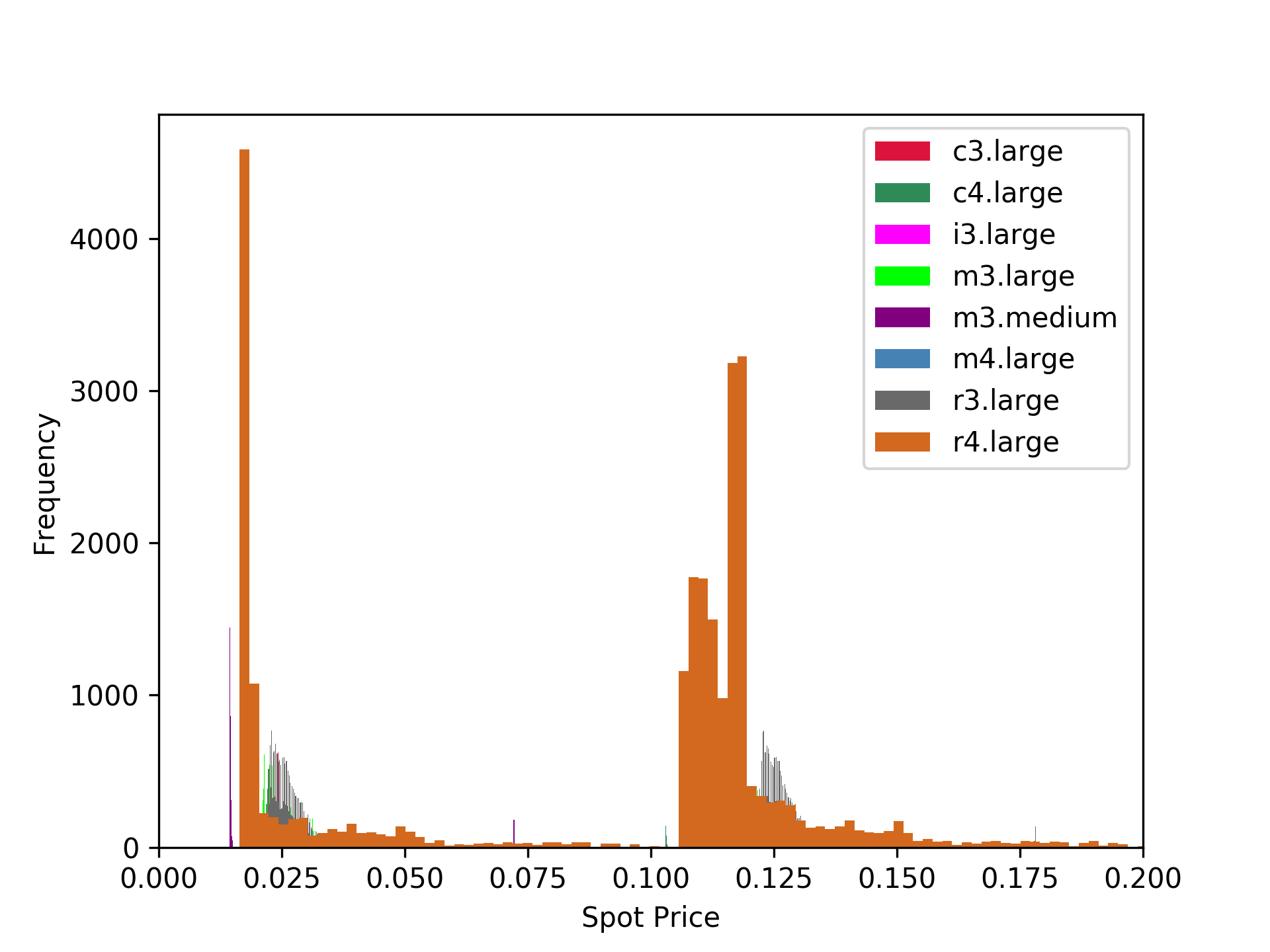} \label{fig:a4}}
\subfigure[eu-west-1c]{\includegraphics[width=0.32\textwidth]{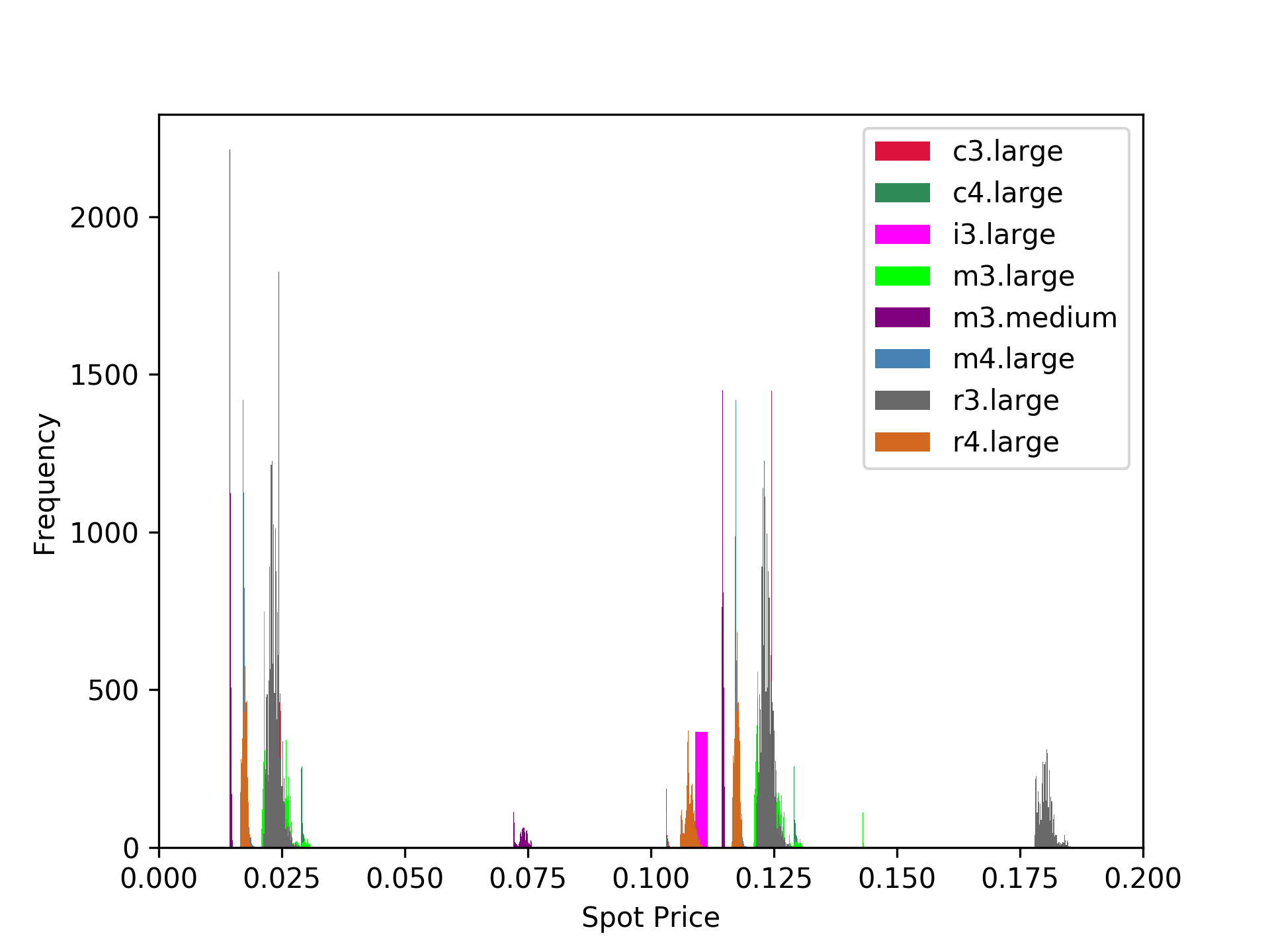} \label{fig:a5}}
\subfigure[eu-west-2a]{\includegraphics[width=0.32\textwidth]{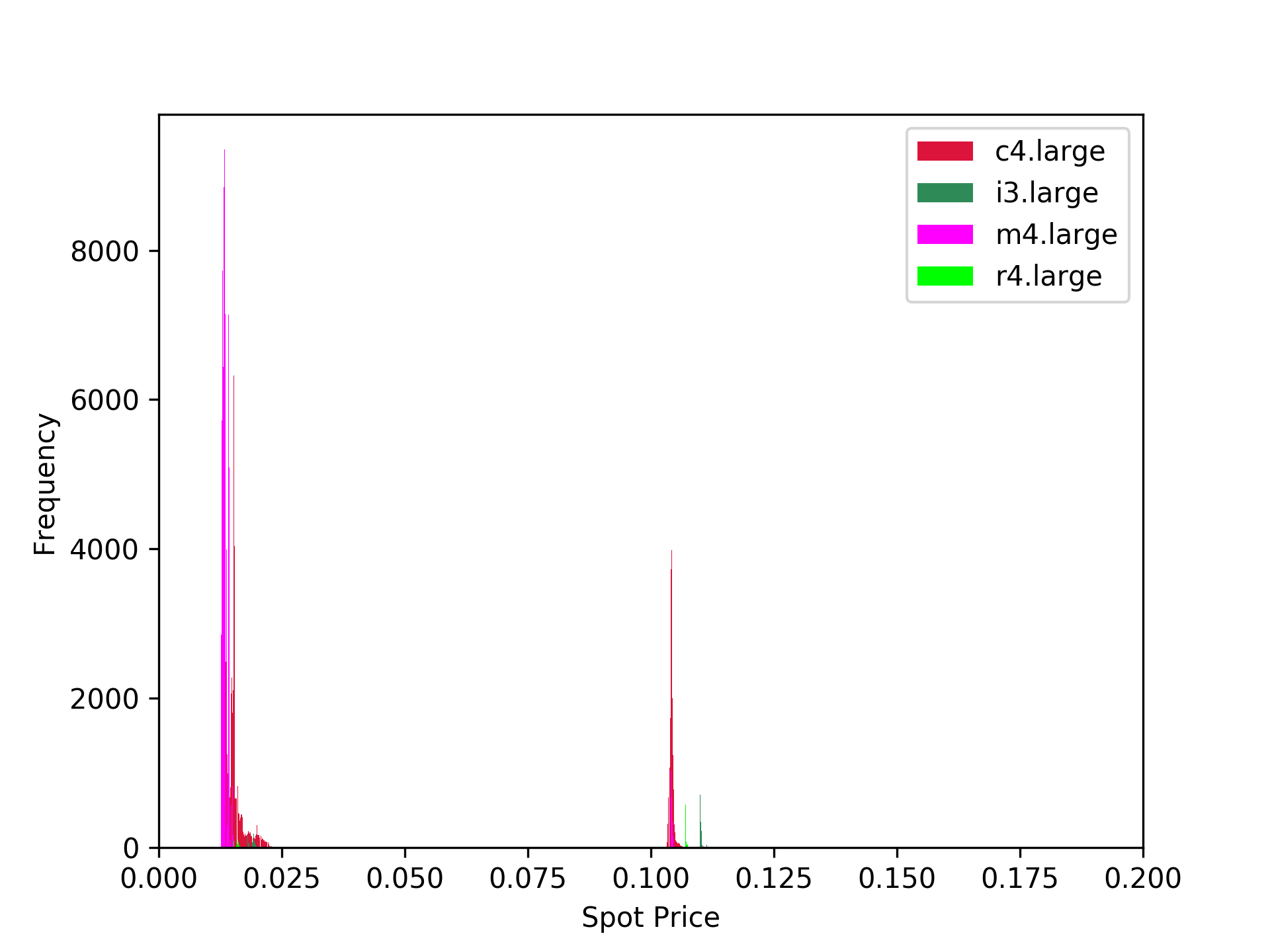} \label{fig:a6}}

\caption{Spot Price Histograms for all AZs within the EU}
\end{center}
 \end{figure*}
 
 \begin{figure*}[!htp]
\begin{center}

\subfigure[us-east-1a]{\includegraphics[width=0.32\textwidth]{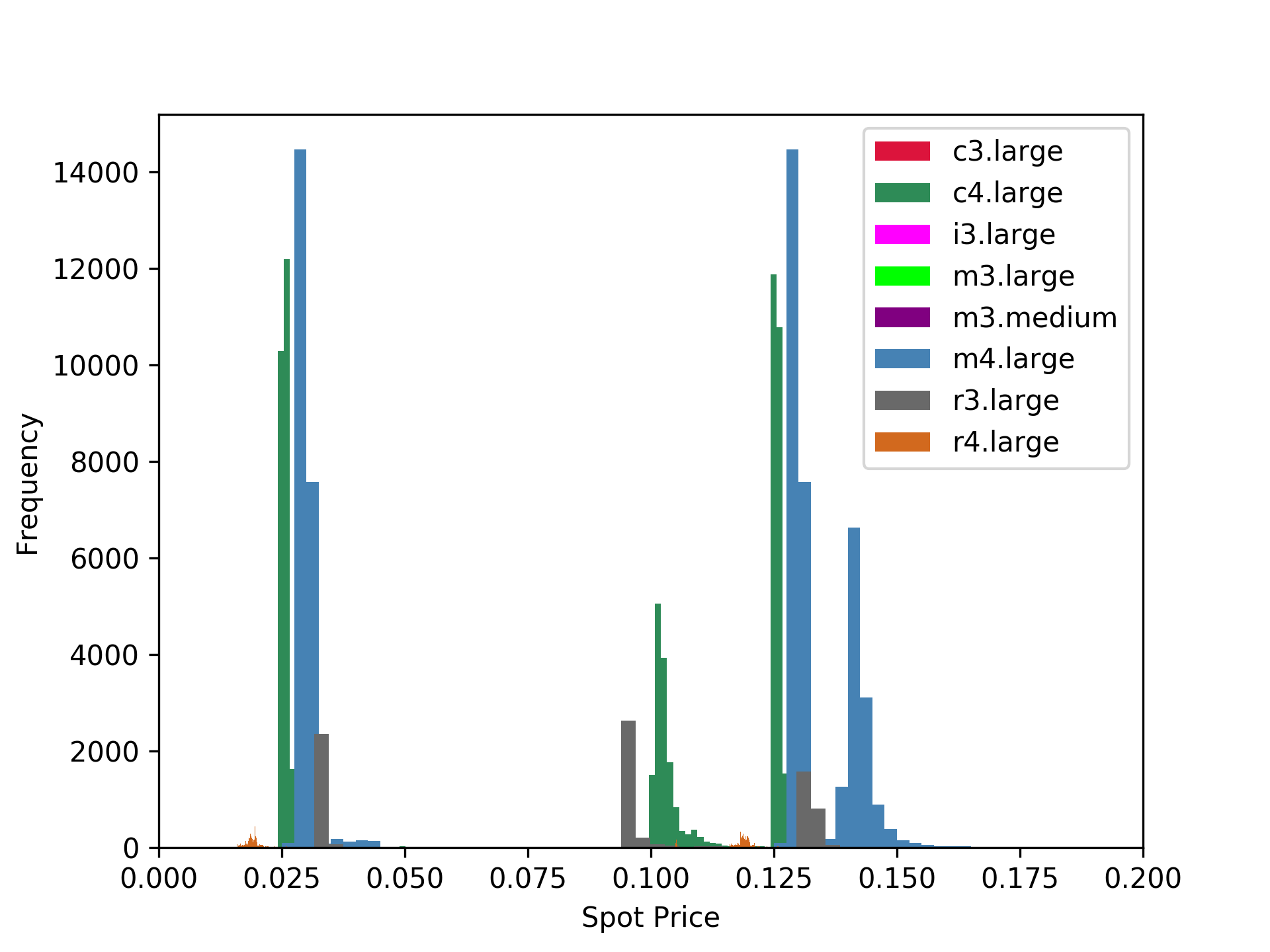} \label{fig:a8}}
\subfigure[us-east-1b]{\includegraphics[width=0.32\textwidth]{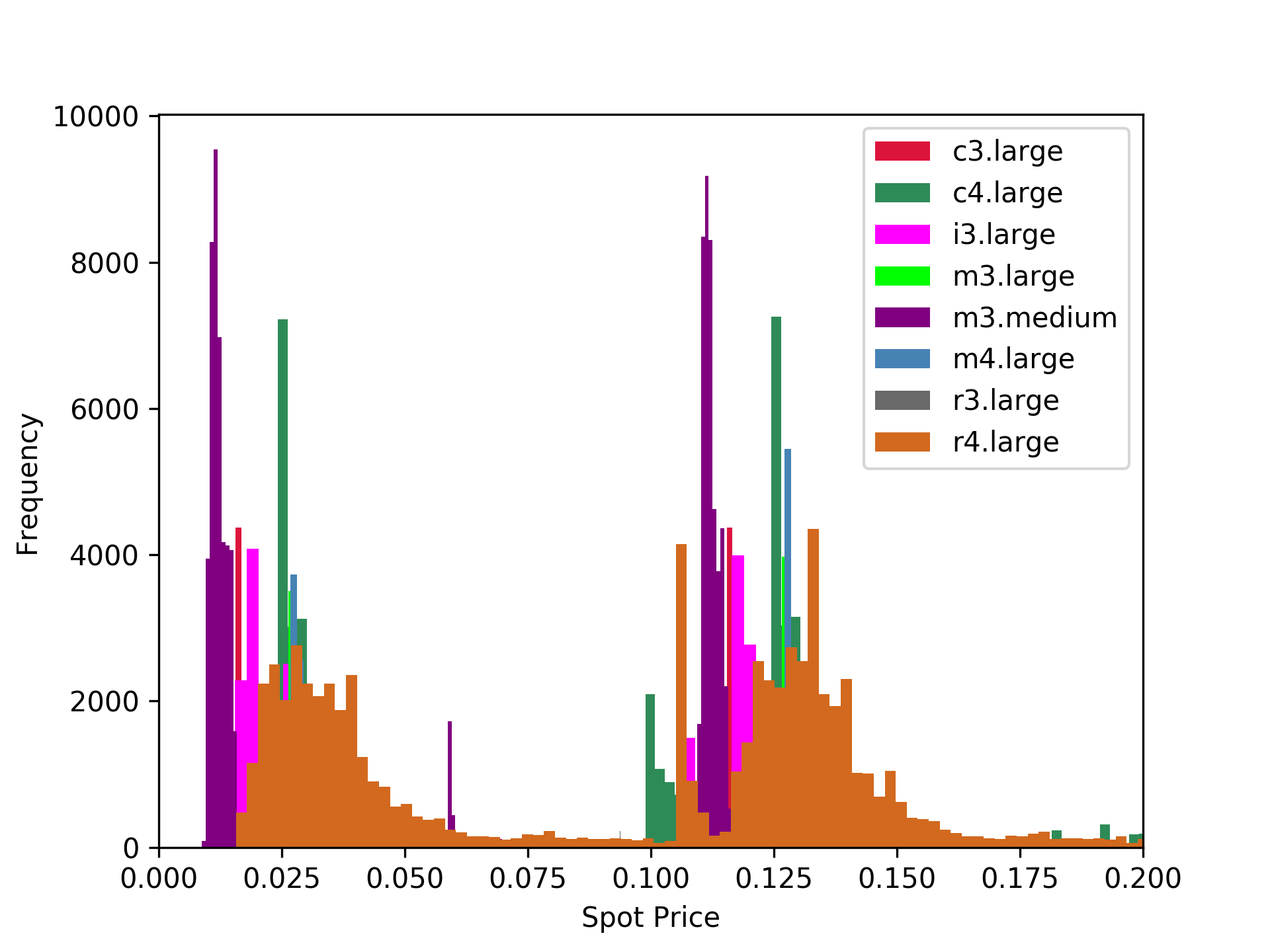} \label{fig:a9}}
\subfigure[us-east-1c]{\includegraphics[width=0.32\textwidth]{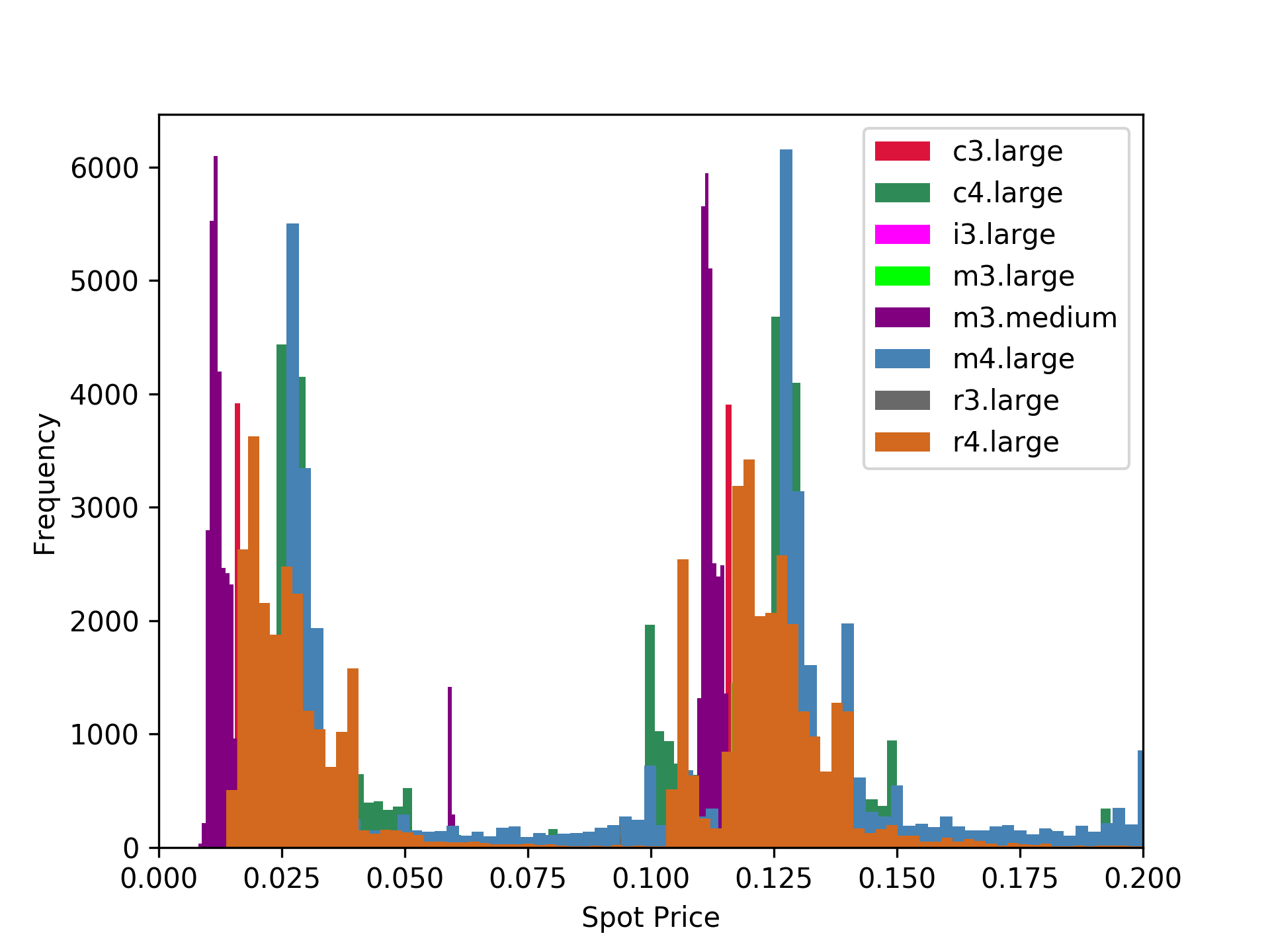} \label{fig:a10}}

\subfigure[us-east-1e]{\includegraphics[width=0.32\textwidth]{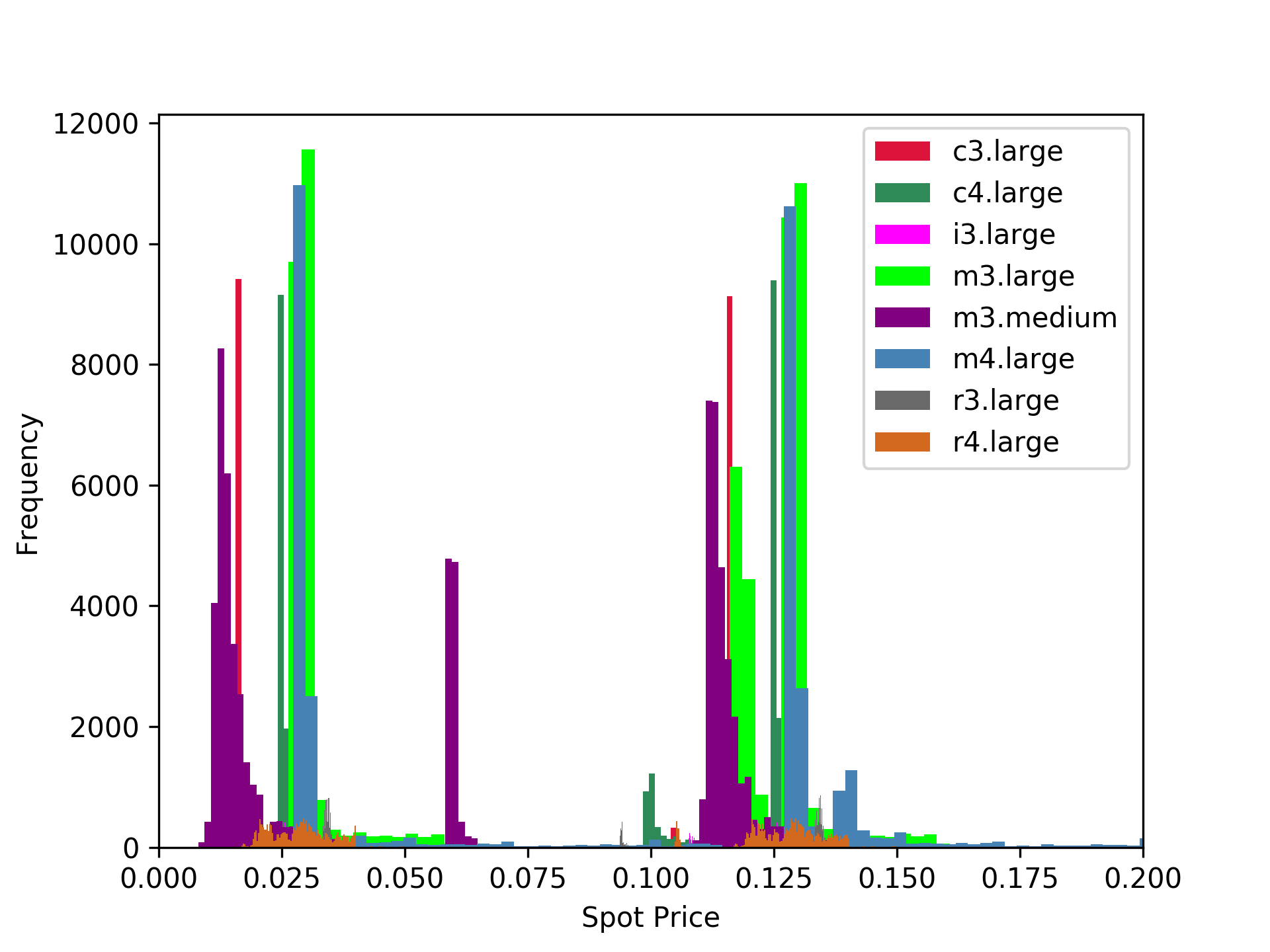} \label{fig:a11}}
\subfigure[us-west-1a]{\includegraphics[width=0.32\textwidth]{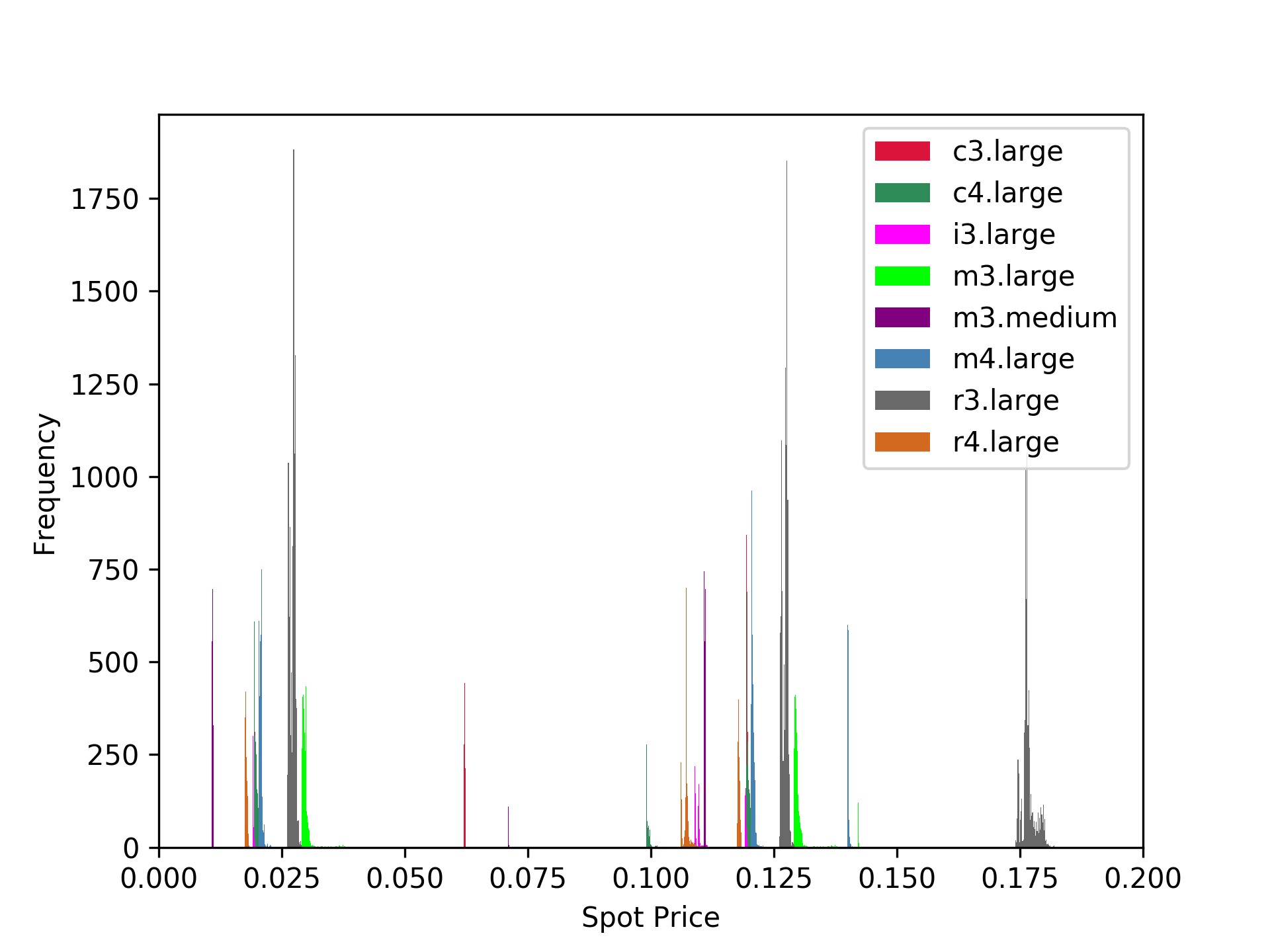} \label{fig:a12}}
\subfigure[us-west-1b]{\includegraphics[width=0.32\textwidth]{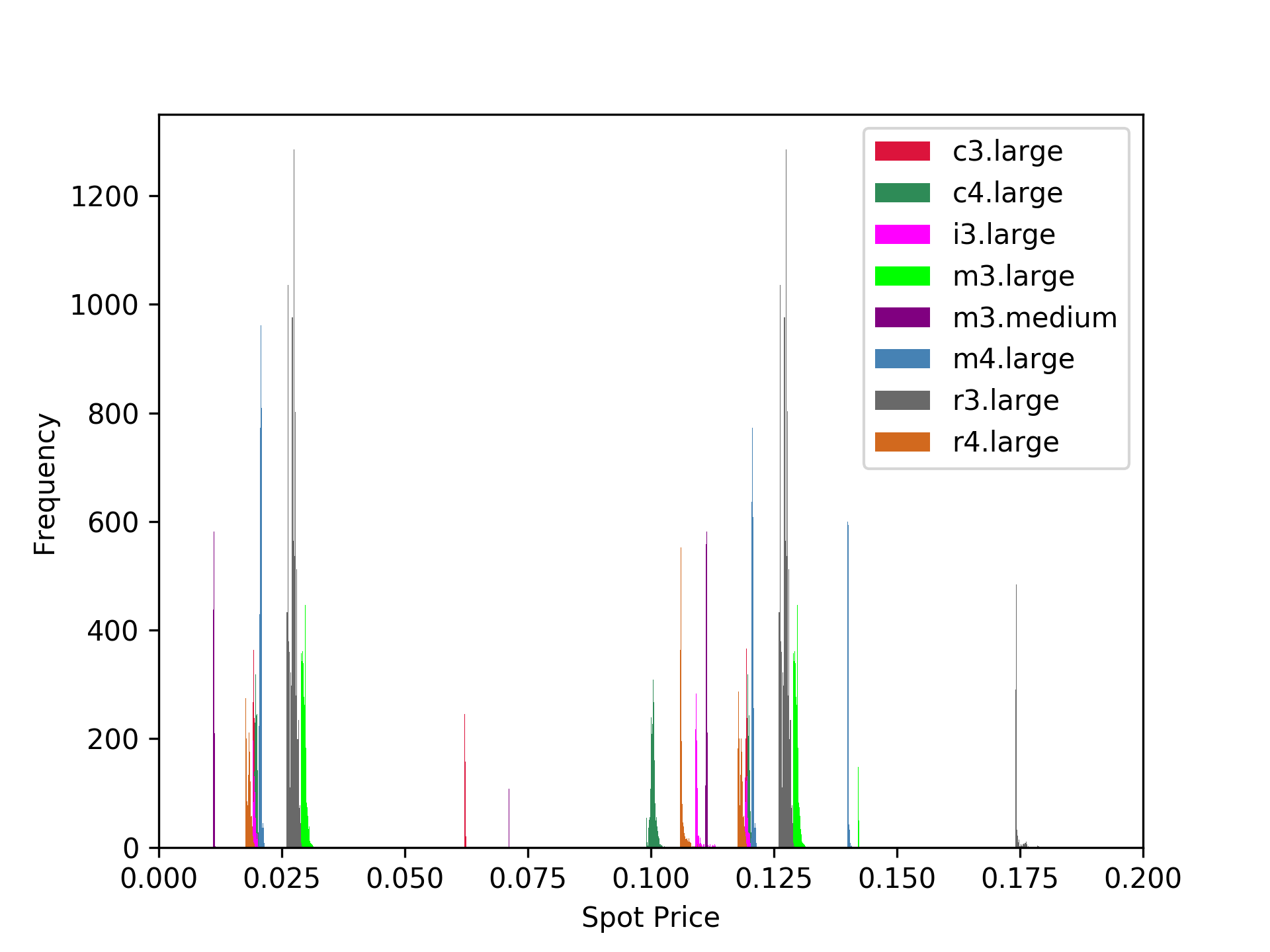} \label{fig:a13}}

\subfigure[us-west-2a]{\includegraphics[width=0.32\textwidth]{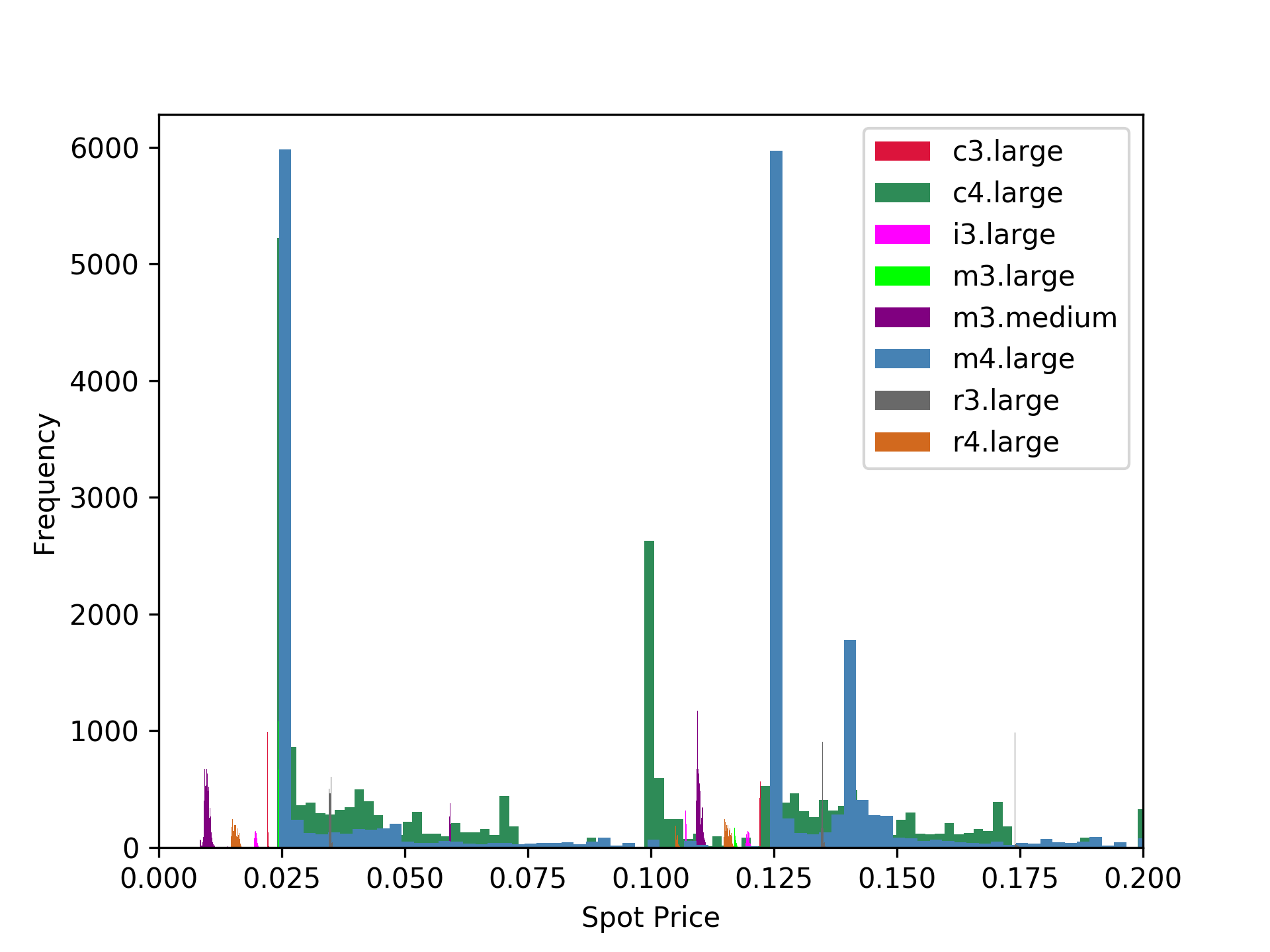} \label{fig:a14}}
\subfigure[us-west-2b]{\includegraphics[width=0.32\textwidth]{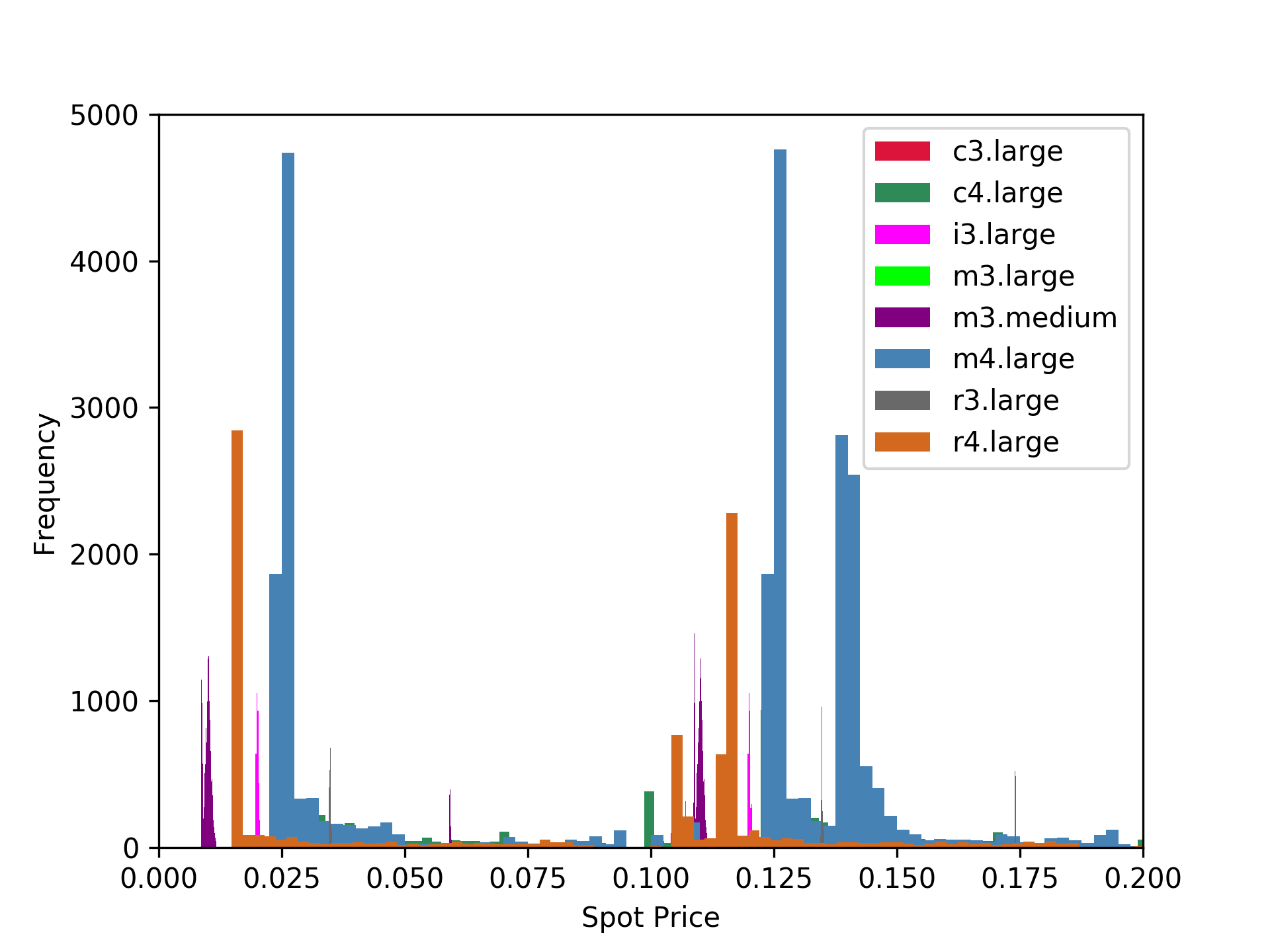} \label{fig:a15}}
\subfigure[us-west-2c]{\includegraphics[width=0.32\textwidth]{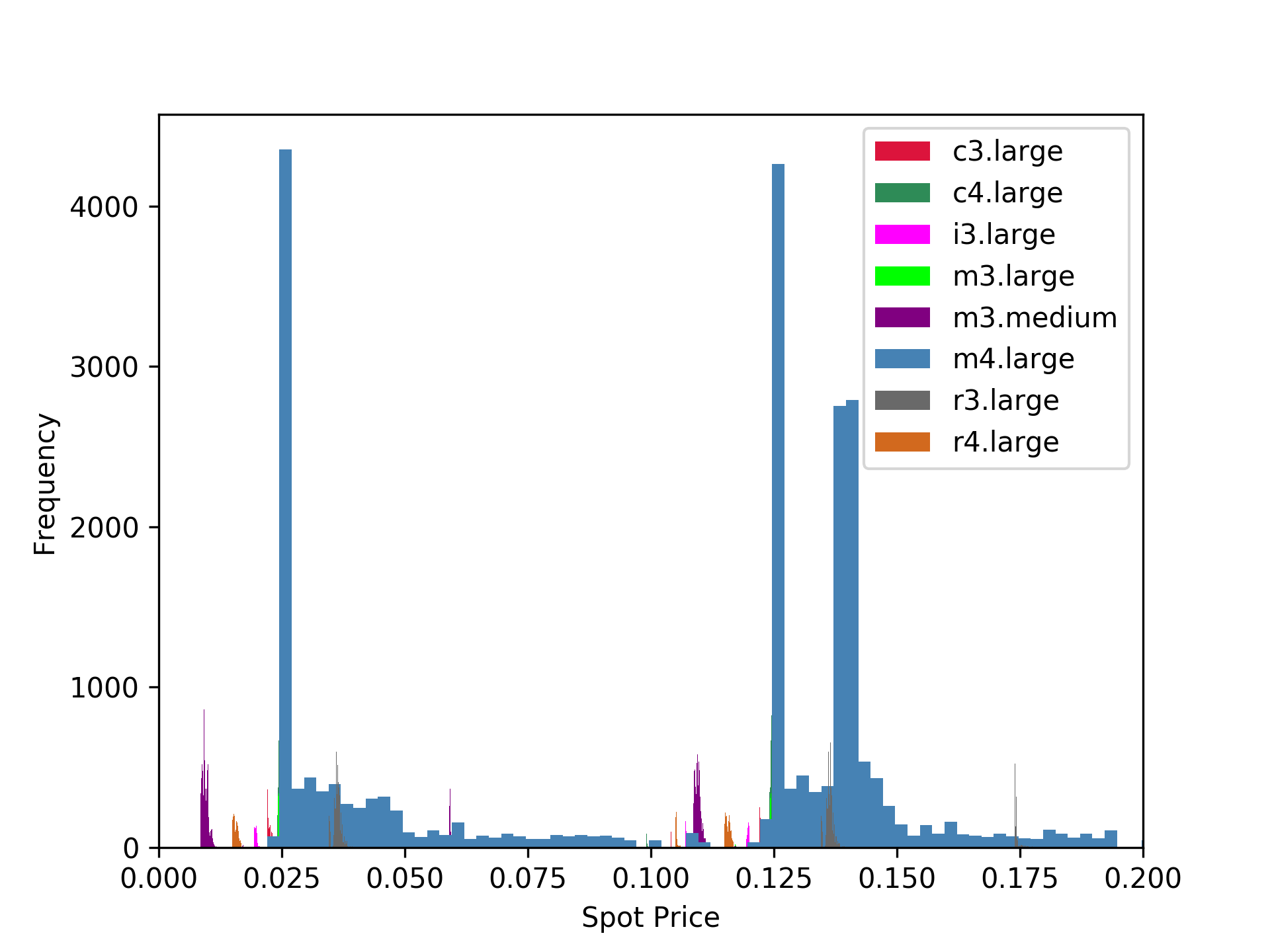} \label{fig:a16}}

\caption{Spot Price Histograms for all AZs within the US}
\end{center}
\end{figure*}

\begin{figure*}[!htp]
\begin{center}

\subfigure[ap-southeast-1a]{\includegraphics[width=0.32\textwidth]{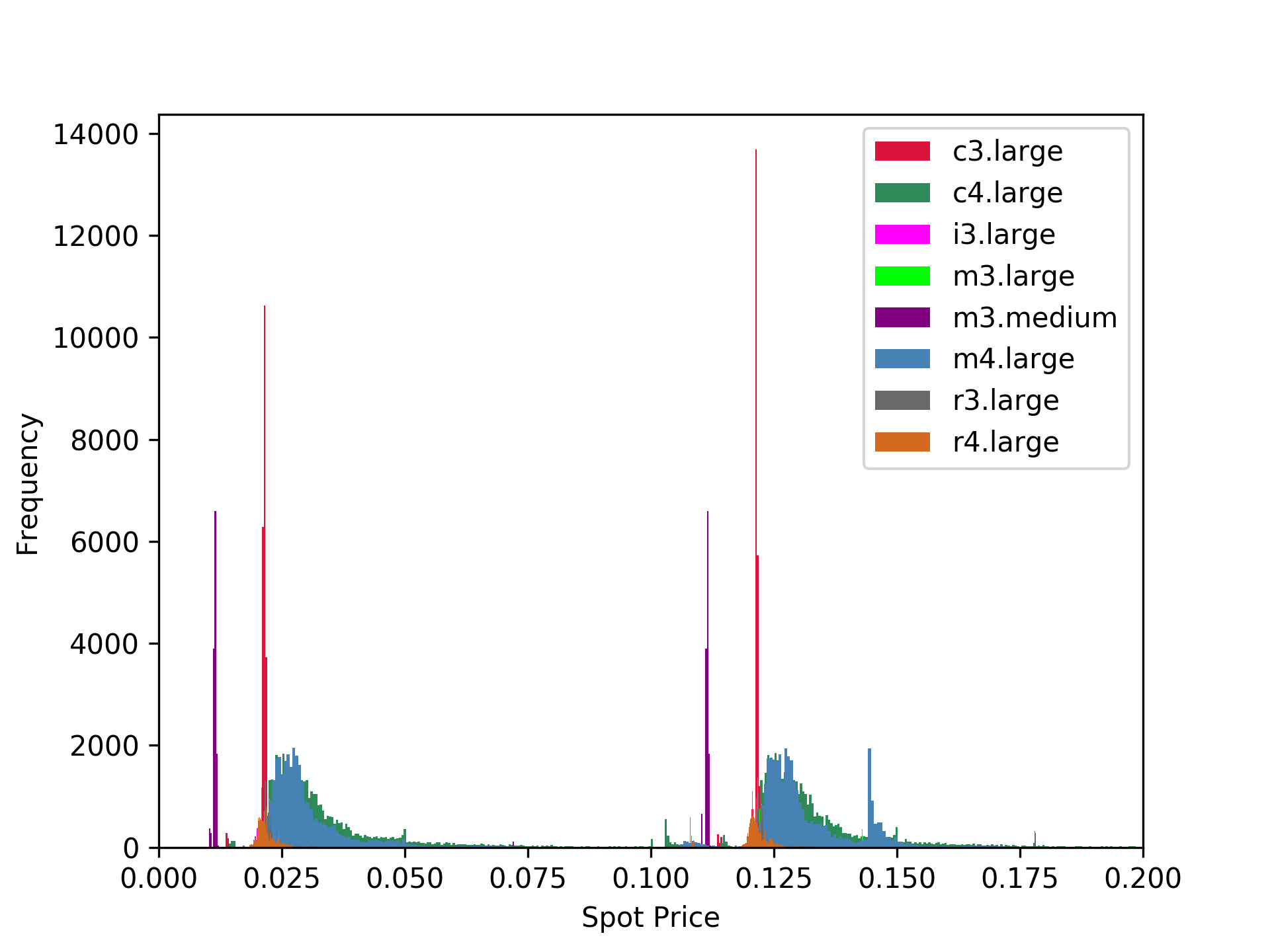} \label{fig:a17}}
\subfigure[ap-southeast-1b]{\includegraphics[width=0.32\textwidth]{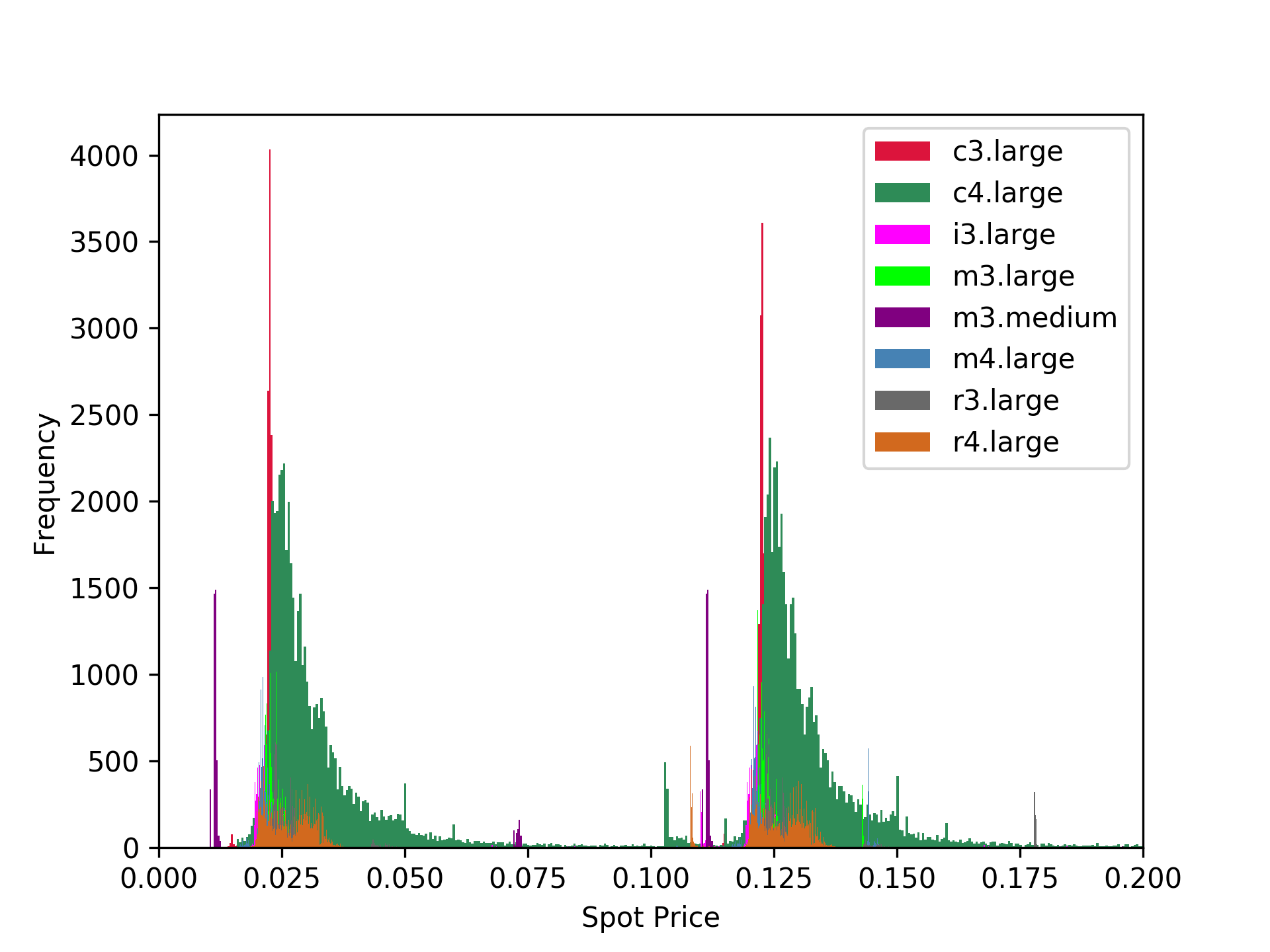} \label{fig:a18}}
\subfigure[ap-southeast-2a]{\includegraphics[width=0.32\textwidth]{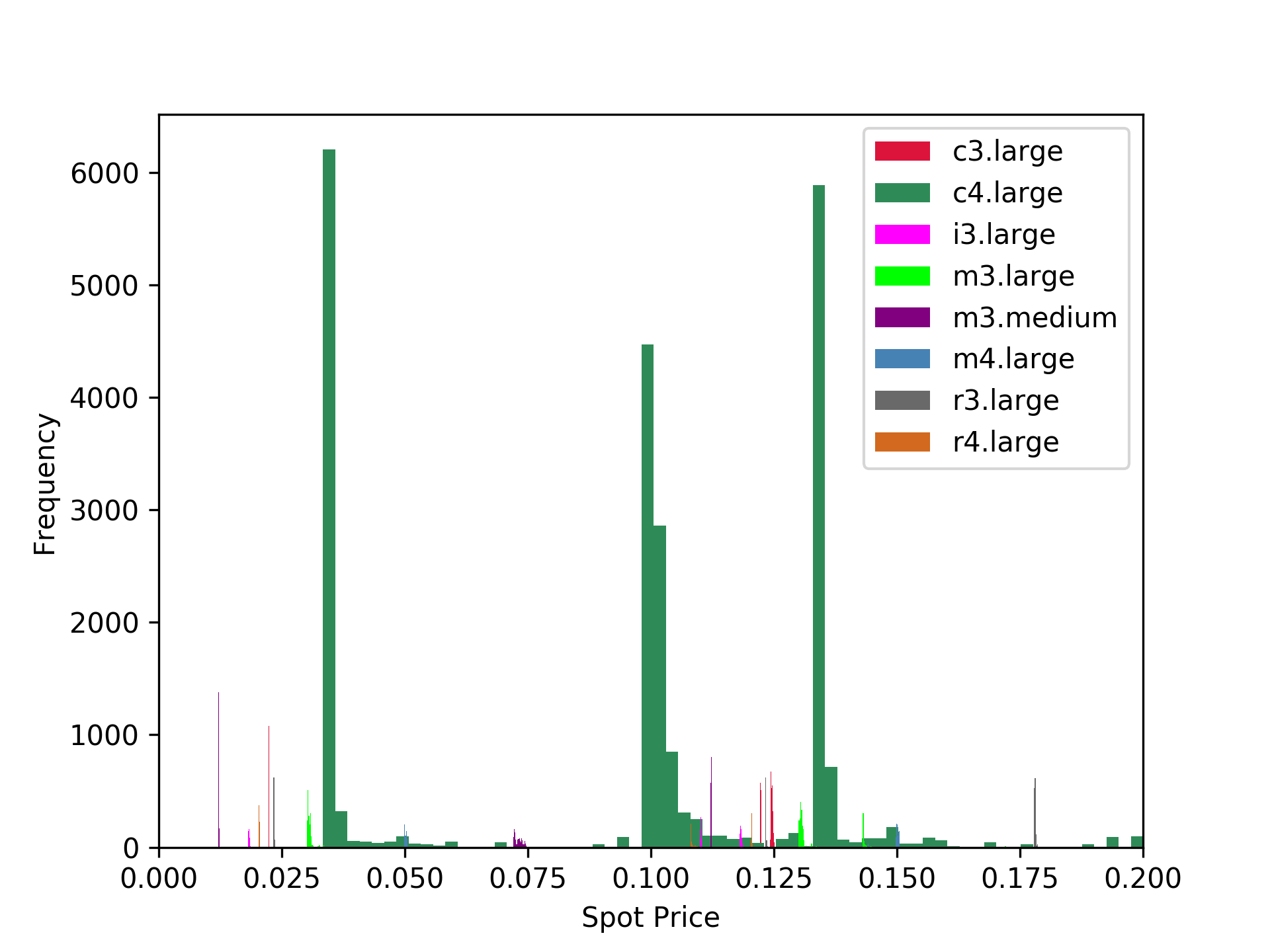} \label{fig:a19}}

\subfigure[ap-southeast-2b]{\includegraphics[width=0.32\textwidth]{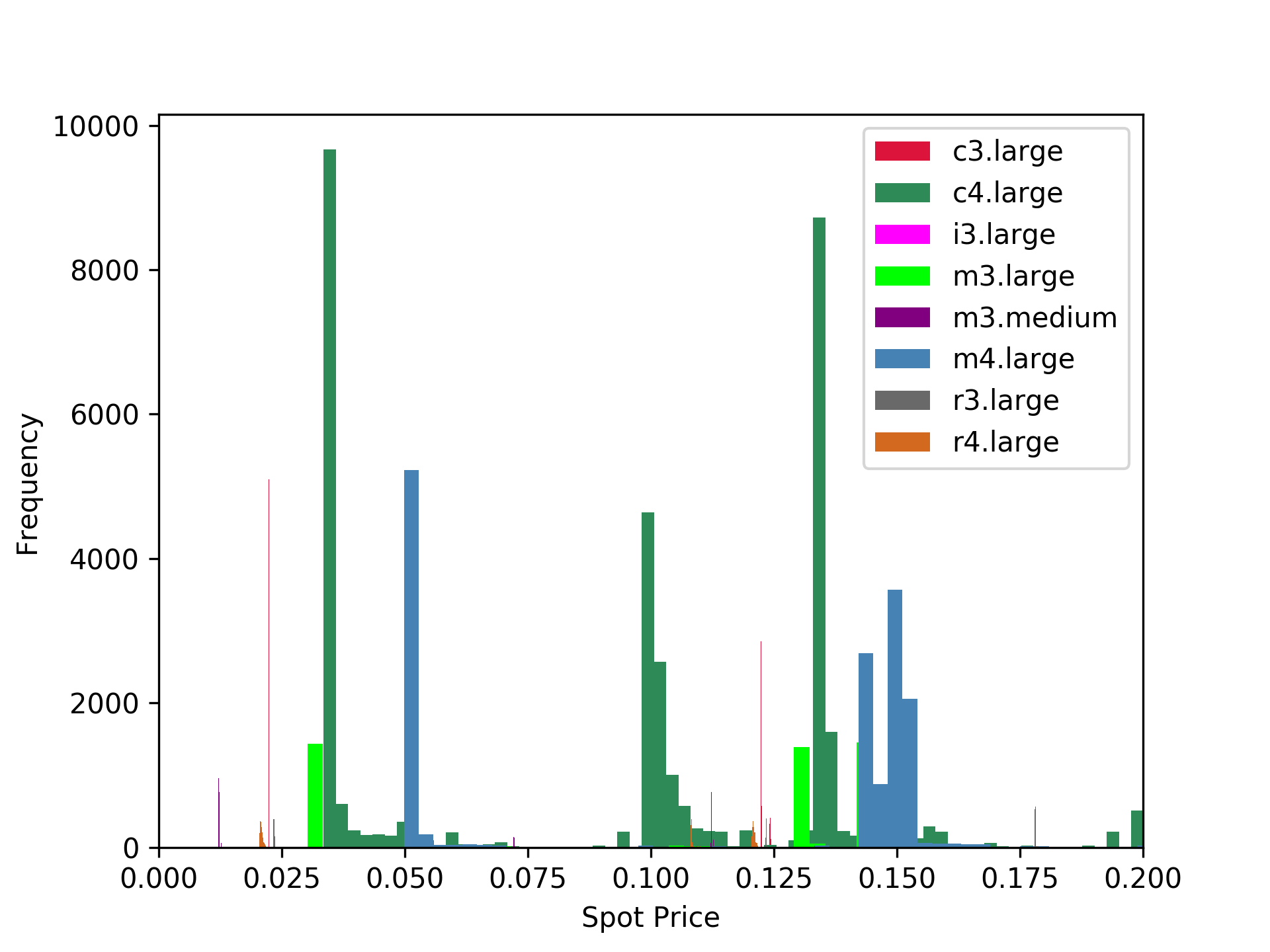} \label{fig:a20}}
\subfigure[ap-southeast-2c]{\includegraphics[width=0.32\textwidth]{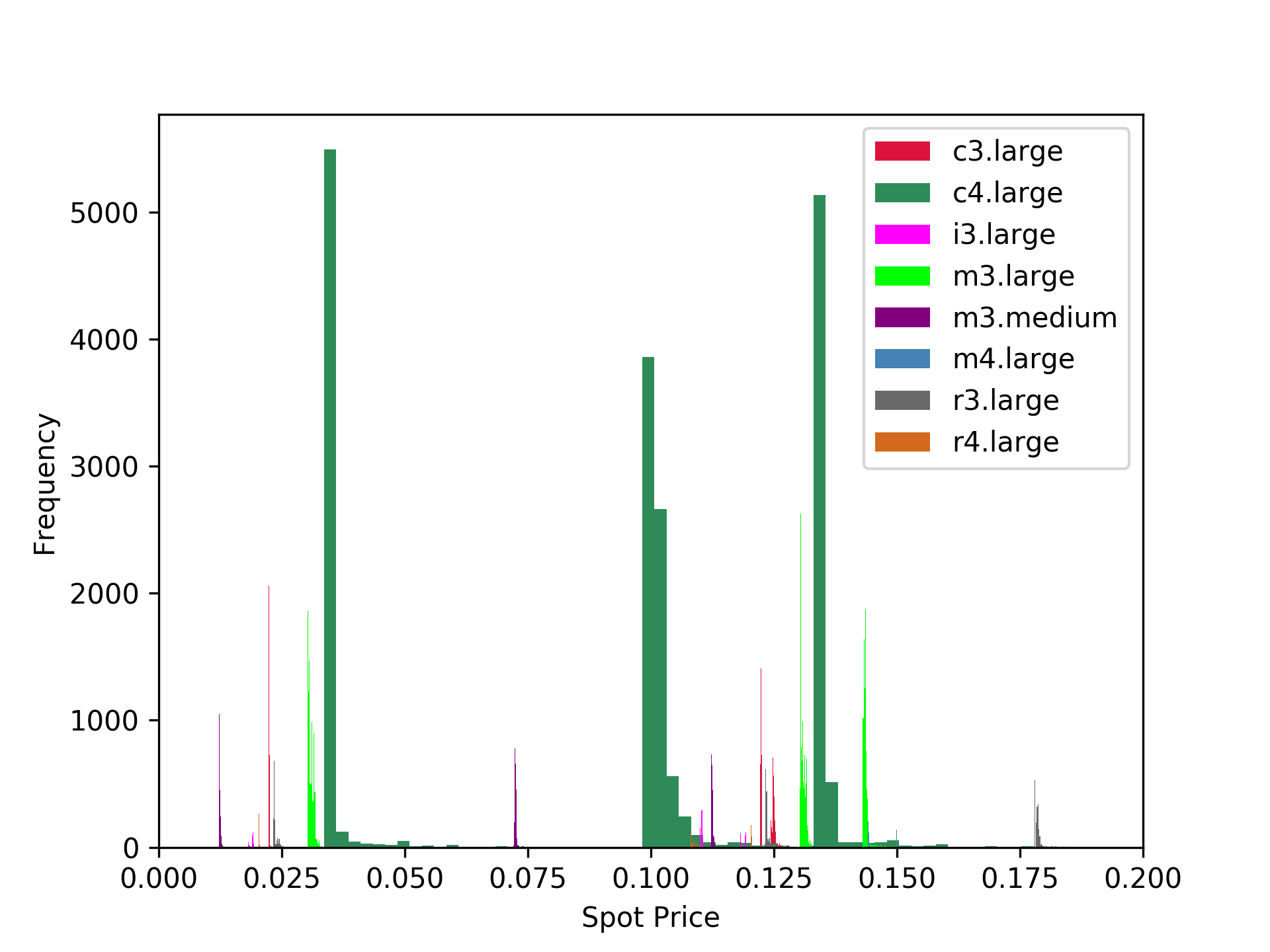} \label{fig:a21}}

\caption{Spot Price Histograms for all AZs within AP}
\end{center}
\end{figure*}

\begin{figure*}[!htb]
\begin{center}

\subfigure[ca-central-1a]{\includegraphics[width=0.32\textwidth]{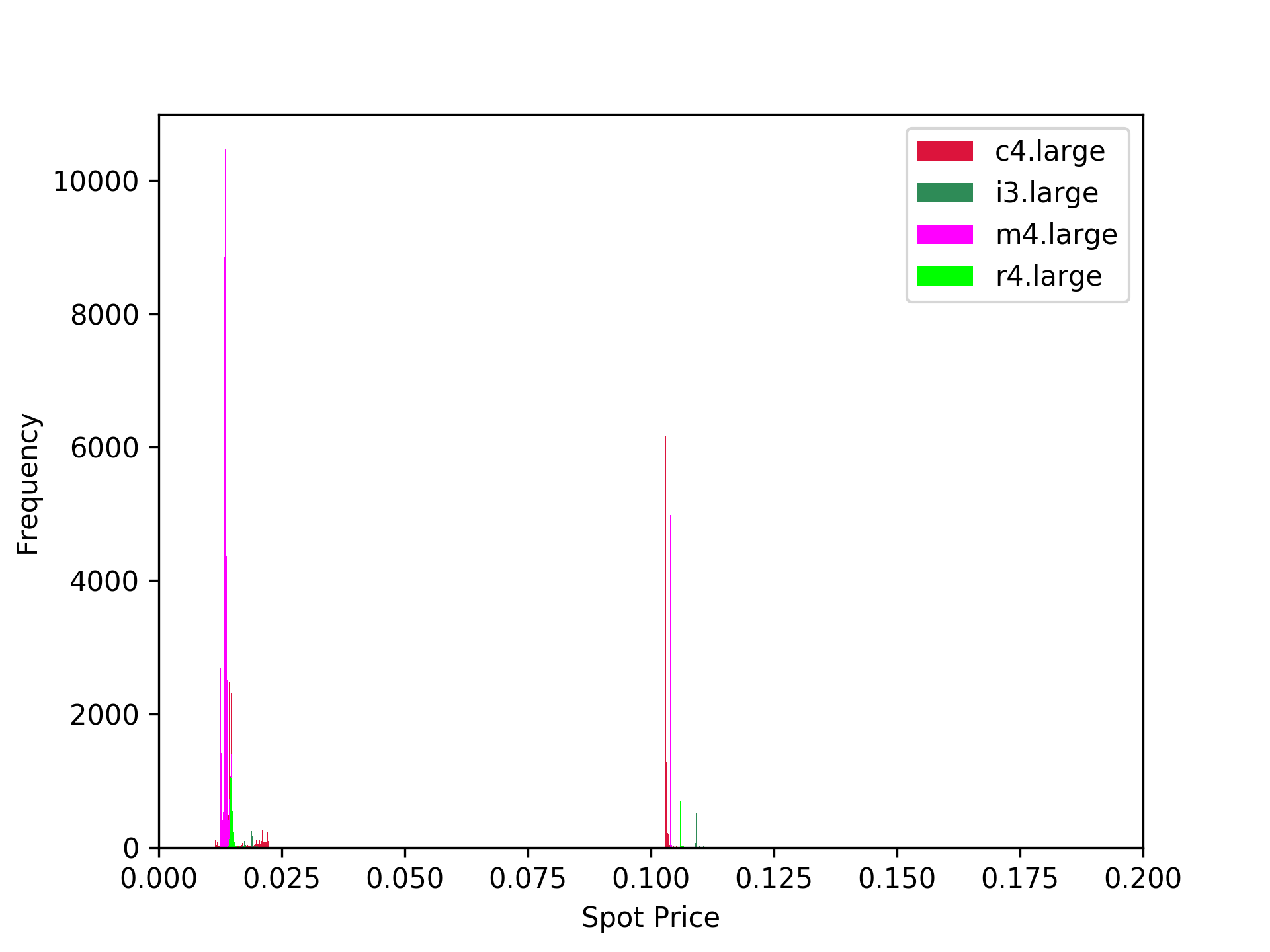} \label{fig:a22}}
\subfigure[ca-central-1b]{\includegraphics[width=0.32\textwidth]{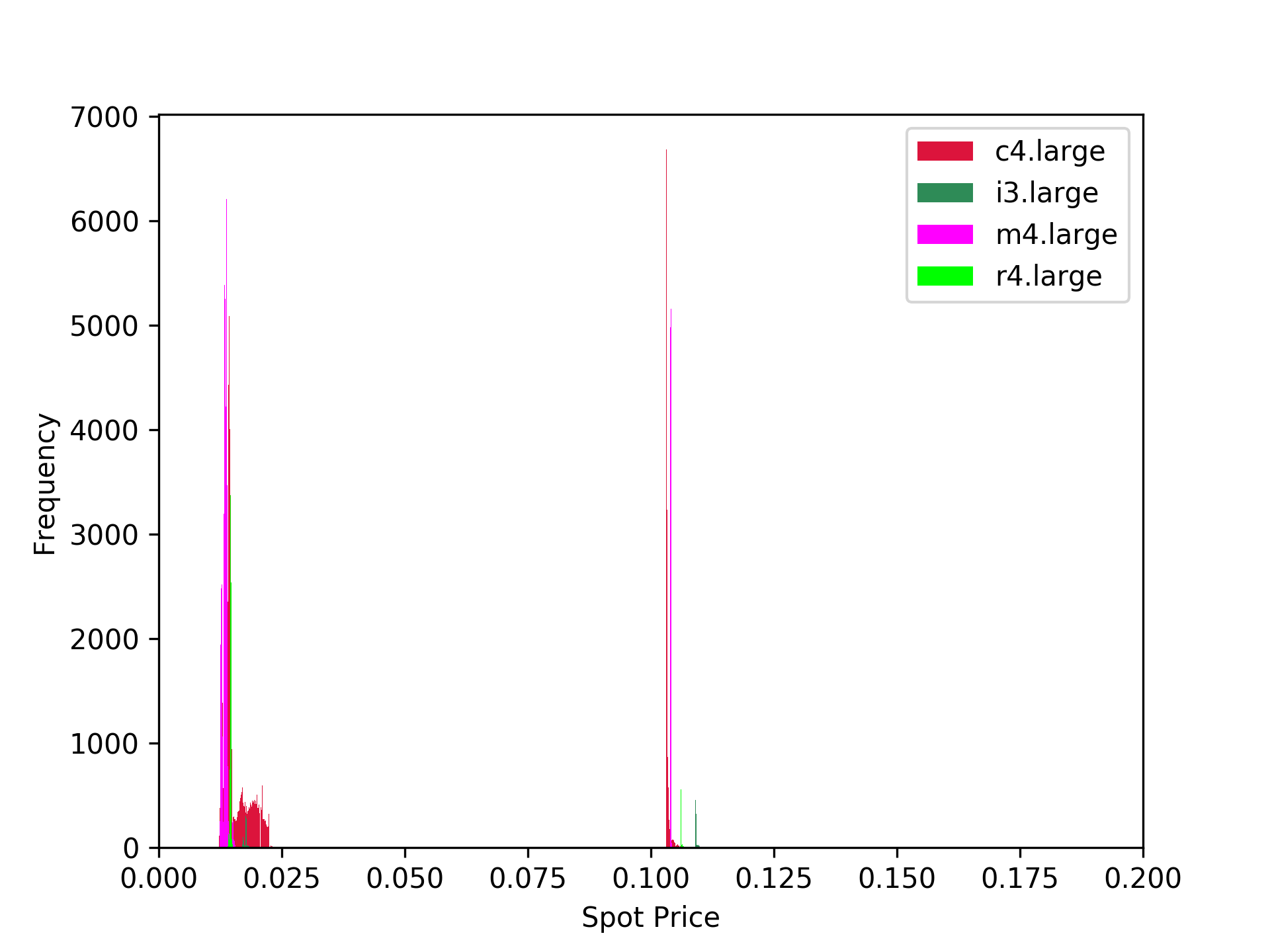} \label{fig:a23}}
\caption{Spot Price Histograms for all AZs within CA}
\end{center}
\end{figure*}

\subsection{Average Price Analysis of All Instance Types within an AWS region}

\subsubsection{EU}

At first glance, the average pricing for the instances seem to be mostly concentrated around the \$0.075 to \$0.080 mark. We can clearly see that \textit{i3.large} instances are very expensive in comparison to the other instances with \textit{m4.large} instances looking like the cheapest overall across the day.

When we related these prices to their standard deviations (Table \ref{table:1}), we noticed some interesting metrics. In general across the EU, prices were very volatile (high levels of standard deviation). This was evident in some instance types more than in others - across the EU, \textit{i3.large}, \textit{\textit{c4.large}} and \textit{m3.large} instances had the three highest levels of standard deviation in the dataset. The three instances that showed the most reliable and stable pricing were the \textit{r4.large}, \textit{m4.large} and \textit{m3.medium} instances (lowest levels of standard deviation). \textit{m3.medium} instances in fact, showed the most consistent and reliable pricing throughout the day, followed by \textit{r4.large} and \textit{m4.large} instances. 

Already, we can begin to see the impact location has on spot instance pricing - \textit{r4.large} instances are some of the most powerful within the AWS offering - however it displays pricing similar to two of the least powerful instances in the dataset. This hourly breakdown offers us a very helpful view of pricing - we can see which hours of the day are cheaper to deploy an instance than in others. For example, we can see that at the beginning of the day (around midnight to 5am and at the tail end of the day - 8pm to midnight) we get some of the cheapest pricing available for \textit{m4.large} instances. \textit{m3.medium} instances show relatively consistent pricing throughout the day, a trend also similarly seen in \textit{r4.large} instances types.

We also performed average pricing analysis over the week - giving us a higher level overview of price differences that we might not necessarily see on a daily basis. As seen in Figure \ref{fig:eu-all-instances-dow}, for \textit{r4.large} and \textit{m4.large} instances, the end of the week seems to be the best time to use them - offering the user the cheapest possible pricing. This, coupled with a low standard deviation allows us to conclude that these are reliable price metrics. \textit{m3.medium} instances' pricing were also very consistent throughout the week. \textit{i3.large} instances were clearly more expensive than all the other instances throughout the week. This, coupled with a very high standard deviation (also seen in \textit{\textit{c4.large}} and \textit{m3.large} instance types) shows the user that the EU is not the best place to deploy such instance types.

\subsubsection{US}

In the United States, we see significant and pronounced price volatility across nearly all the instance types. 
We see both very high levels of standard deviations (Table \ref{table:2}) as well as high average price metrics throughout the day and throughout the week. 
Only two instance types (as seen in Figure \ref{fig:us-all-instances-per-hour}) display significantly low levels of standard deviations throughout the entire time period - \textit{m3.medium} and \textit{r3.large} instances. \textit{m4.large} instance types display the most volatile pricing in the region, closely followed by \textit{i3.large} and \textit{m3.large} instance types. We see a few interesting things here. First, the type of instance clearly does not have as large an impact on price as location does. \textit{m3} instance types are the least powerful in our dataset, yet we're seeing both very high and volatile pricing across the board. Conversely, we see reliable and low price metrics reported for \textit{r3.large} instances, some of the most powerful in our dataset. We also see this replicated again in the week.

\subsubsection{Asia-Pacific}

The Asia-Pacific region, in comparison to the EU and US regions show both low price points coupled with low standard deviations (Table \ref{table:3}). 
We see very low and reliable pricing in the \textit{m3.medium}, \textit{r4.large}, \textit{\textit{c3.large}} instances (Figure \ref{fig:ap-all-instances-per-hour}). We again see the same pattern we saw earlier - more powerful instance types appearing at very low and reliable price points. In general, the more powerful the instance type, the more favourable the pricing in the Asia-Pacific region. \textit{i3.large} instances once again display the most volatile pricing across the board, but for the most part, our previous conclusion holds. When we further examined some of the price metrics, we saw a sizeable difference in pricing with the Asia-Pacific region and the EU and US: \textit{r3.large} and \textit{r4.large} instances have max prices of \$0.1934 and \$0.14 per hour. In the EU, we see max prices 17x and 18x that for \textit{r3.large} and \textit{r4.large} instance types respectively, and 15x that in the US for both instance types.
Across the week (Figure \ref{fig:ap-all-instances-dow}), we see also see these low prices reflected.

\subsubsection{Canada}

The Canada region hosts the least amount of instance types from our original selection - only the more powerful \textit{\textit{c4.large}}, \textit{m4.large}, \textit{i3.large} and \textit{r4.large} instance types are present.

From these graphs (Figure \ref{fig:ca-all-instances-per-hour} and Figure \ref{fig:ca-all-instances-per-dow}) the Canada region exhibits some of the cheapest pricing on average per hour and across the week for the different instance types. For the \textit{i3.large} instance type, we see mean prices at around \$0.07 per hour for the instance with the lowest standard deviations recorded across all four regions at \$0.045 (Table \ref{table:4}). In the Asia Pacific region, mean prices are at \$0.087 with standard deviations of \$0.167. In the US, \$0.081 with \$0.115 as the standard deviation and in the EU, \$0.181 with a standard deviation of \$0.496. From this we can clearly see that it would be much cheaper to deploy an \textit{i3.large} instance in the Canada region than in any of the others. 
We can also apply the above analysis to all the other instance types.

We can therefore conclude, that the more powerful the instance type, the cheaper and more reliable it is to deploy in Canada.

\subsection{Average Price Analysis of Each Instance for every AWS region}

This analysis plotted every single instance type's pricing for every single AWS region. This allowed us to perform cross-regional comparisons and offered another view of the price data we had gathered. For some of the regions, some instance types were not deployed and as a result do not appear on some graphs. 

We see some very clear patterns straight away. In any instance type that is deployed in Canada, we see both its lowest and most reliable price metrics. If we take a look at Figures \ref{fig:each-instance-every-region-per-hour-average-pricing} and \ref{fig:each-instance-every-region-per-dow-average-pricing}, we see this very clearly. This also serves to reinforces the conclusions we came to earlier. The less powerful the instance, the more reliable it is to deploy in the EU or US - in the ``m'' instance types we generally see cheaper pricing in those two regions.

For example, if one was to deploy an \textit{m4.large} instance and had the choice between the AP region and the EU, one would obtain much more reliable pricing in the EU with a lower mean price and standard deviation than in AP. There are however, some anomalies - for example \textit{m4.large} instance types being amongst the cheapest to run in the EU but the most expensive in the US. What one can see however, is that with these different price metrics we can begin to make informed decisions about where to deploy an instance. If (for instance) there were certain constraints that meant we could only deploy in a specific region over another, we can use this analysis to make the best choice for a particular use case.

\subsection{Spot Price Histogram Price Frequency Analysis}

\subsubsection{EU Region Analysis}

We can see that \textit{m3.medium} instances are generally the cheapest to use amongst the
availability zones (AZs) in which they have been deployed. However, in some AZs we
do not see as high a frequency of data for \textit{m3.medium} instances as some of the others. 
Note that the \textit{eu-west-2a} AZ did not record any data for the \textit{m3.medium} instances.

In the \textit{eu-central-1a} AZ (Figure \ref{fig:a1}), it was generally cheaper to deploy on a \textit{c4.large}
instance than it was on an \textit{m3.large} instance. This is particularly interesting as \emph{c4} instances are ``optimized for compute-intensive workloads" \cite{amazonawsinstancetypes}. c4 instances are also more powerful than both \textit{m3} and \textit{m4} instances. This is a trend that was only really
noticeable in the \textit{eu-central-1a} AZ due to the amount of pricing data generated for
\textit{\textit{c4.large}} instances. Looking across the other AZs however, \textit{\textit{c4.large}} instances had
in several cases much cheaper pricing per hour than the \textit{m3} and \textit{m4} instance types.

Towards the other end of the scale, \textit{r3} and \textit{r4} instance types displayed some
unexpected pricing trends. \textit{r3} and \textit{r4} instance types are amongst the most powerful on
AWS and are used to run ``high performance databases...distributed web scale in-memory caches" \cite{amazonawsinstancetypes} etc. However, in the \textit{eu-west-1b} AZ (Figure \ref{fig:a4}) we see \textit{r4.large} instances displaying modal prices of around \$0.02 p/hour with a frequency of over 4000
and then of around \$0.12 p/hour (with a frequency of just over 3000). \textit{r3.large} instances also generally fall into these above two ``price brackets" — (Figure \ref{fig:a3} - \textit{eu-west-1a} and Figure \ref{fig:a5} - \textit{eu-west-1c} AZs).

This is very interesting as it shows that 1) location of
deployment is extremely important (specifically AZ) and 2) all these analyses must be
taken together in context to find the best deployment area. A possible scenario - let’s
say a developer wanted to run a workload on a ``r'' instance type but data protection
rules limited them to running this workload in Europe. From this analysis, the developer
could run this workload quite comfortably in the \textit{eu-west-1b} region and get prices
comparable to Canada or Asia Pacific regions.

\subsubsection{US Region Analysis}

\textit{r4.large} instance types appear regularly on the left most side of the chart (i.e. the
least expensive). In the \textit{us-west-2b} AZ (Figure \ref{fig:a15}), we see that \textit{r4.large} instances are cheaper
than \textit{m4.large} instances with prices at around \$0.02 per hour even though \textit{r4.large}
instances are a lot more powerful. We see this again in the \textit{us-east-1c} AZ (Figure \ref{fig:a10}) with
the mode pricing being around \$0.02-\$0.03 per hour as well as \$0.015-\$0.125.

\textit{m4.large} instances follow their own distinct pattern across all the different AZs. Their
two highest (modal) pricing points are either \$0.025 or \$0.125 per hour. We can see this
most clearly in the \textit{us-west-2a}, \textit{2b} and \textit{2c} AZs. In the \textit{us-west-2a} AZ (Figure \ref{fig:a14}) the two modal
price points occur 6000 times with the next closest price point occurring around 1800 times.

In all AZs where they've been deployed, \textit{m3.medium} instance types are consistently the
cheapest instance type to use.

\subsubsection{AP Region Analysis}

\textit{c3.large} instances show modal prices of either \$0.025 or \$0.125
an hour. The \textit{c4.large} instance types generally fell within 3 modal pricing points in the
\emph{ap-southeast-2} regions - either \$0.03, \$0.10 or \$0.14 an hour.

The Asia Pacific region appears to be more popular with larger and more powerful
instance types as that is where most of the data points are recorded for.
Smaller and more General Purpose instance types have much smaller data frequencies
– although where they do occur, the data points generally show cheap pricing. From this
graph, we can clearly see advantages with deploying more powerful instance types
(mainly c types) in this region than in the EU or US.

\subsubsection{Canada Region Analysis}

m4.large instance types are the cheapest instances to use in both the \textit{ca-central-1a} and \textit{ca-central-1b} AZs. In general prices across the zone are very inexpensive
with prices not going past \$0.11 an hour mark. They also occur with high frequency. If
one was deploying one of these 4 \textit{.large} instances, one would deploy them here than
any of the other regions. 

Also, this graph has very few data points meaning that the region is not widely used — a factor one could take advantage of to eliminate
application ``hotspots" as well as to get the lowest prices available.

\section{CROSS-COMPARISONS}

\subsection{Average Price Analysis}

The average price analysis over the day and week show us some interesting insights.
We can see clear price differences in instance types as we go from region to region. In
the Asia-Pacific and Canada regions we generally see much lower and more reliable
pricing than in the EU and US regions. However, this is also instance type dependent.
Overall, \textit{m3.medium} instance types consistently show the most reliable pricing across
all of the regions they are deployed in while \emph{i3.large} instances show the most
volatile. An interesting point to note is that \textit{m3.medium} instances' mean prices are
around the same point across all regions in which they are deployed - \$0.063 per
hour.

We can use this analysis to examine which regions offer best value for the instance
types we require. If one was to deploy an \textit{i3.large} instance type, one of the most
powerful in our list of instance types, we'd would be looking at highly unstable pricing in the
EU, US and Asia Pacific regions. However, in Canada, we'd would be obtaining very cheap
and reliable pricing per hour.

If one wanted to use a quite powerful \textit{c3.large} instance, it would be much cheaper to
deploy in the Asia Pacific region with a mean price of \$0.0764 per hour, a low standard
deviation of 0.05, and max price of \$0.364 than in the EU or US. If one wanted to deploy a less powerful \textit{m4.large} instance, however, one would much rather deploy it in the EU where the prices are on average \$0.052 p/hour with a standard
deviation of 0.054 in comparison to \$0.089 p/hour and a standard deviation of 0.067 in the
Asia Pacific region.

Based on all these analyses, there is a clear pattern identified. Overall, the more powerful
the instance, the cheaper and more reliable the pricing one would get in Canada than in
any of the other three regions. This is closely followed by the Asia Pacific region. On the
other hand, the less powerful the instance the cheaper and more reliable (in terms of
pricing) it is to deploy in the EU and US regions.

\subsection{Spot Price Histograms}

The Spot Price Histograms show us the frequency of prices for each instance type
across the AZ. In the EU and US regions, we see a pattern - the ``m" instances are generally the
cheapest to run and appear in high frequencies towards the leftmost side of the graph.

However, in some AZs within these regions, we see that the more powerful instance
types are among the cheapest to run - \textit{r4.large} instances being an example.
In the Asia Pacific and Canada regions, the \emph{.large} instances present a double
advantage of lower demand and lower prices.

These histograms can allow a developer to take a range of price points for a
particular instance. A developer can then utilise the likelihood (modality) of those price
points occurring and make a confident max bid price.

\section{RELATED WORK}

There has been a significant body of research done into Amazon EC2 Spot Instances. However, most of the research done has been mainly focused on trying to determine how the pricing for the instances is generated. Bodies of work such as \cite{artur-andrzejak, ben-yehuda, liang-zheng} look at modelling the
pricing strategy for Spot Instances as well as determining the best bid prices to make on an instance. Papers \cite{artur-andrzejak, liang-zheng} try to find the lowest price possible for a user to spend,
whilst maintaining the highest availability possible for the instance.

Other bodies of research, more related to this paper, provide more meaningful insights. Wang et al., in \cite{pennsylvania-state-university-website} followed a more empirical approach to analysing Spot Instances, relating it to finding a more cost-effective procurement of compute resources for customers. They identified four key features a potential tenant of a Spot Instance should examine, ``lifetime of an instance average spot price during lifetime simultaneous revocations, and startup delay", created a quantitative model, and then used this model to develop ``computationally-efficient predictors" of pricing \cite{pennsylvania-state-university-website}.

Their conclusions revealed that ``the evolution of spot price depended on a spatially coarse (e.g., data center or availability zone wide) load metric" \cite{pennsylvania-state-university-website}. Their conclusions tie in fully with our own that the data center region or AZ had a significant impact on price.
Approaching Spot Instances from an alternate angle, Li et al., recently in \cite{li-zheng} performed a comparative investigation between Spot Instances and fixed-price instances using a Systematic Literature Review.  They highlighted the fact that cloud consumers were concerned at ``unpredictably frequent interruptions when using spot services" \cite{li-zheng} - something that was putting them off utilising the spot market fully. The ``academic community strongly advocate[d] the Cloud spot market" \cite{li-zheng} but that that
was still not being wholly reflected by consumers.

In recent times, however, this has started to change. Spotinst \cite{spotinstwebsite}, a cloud company, launched in 2015 to ``Reduce 50-80\%'' of cloud computing costs for companies on EC2 via the use of Spot Instances.
It does this by utilising predictive algorithms to ``to predict Spot behaviour, capacity trends, pricing, and interruption rate'' \cite{spotinstelastigroup}, rebalancing whenever there is risk of an interruption, and falling back
onto on-demand instances when Spot Instances are not available to use.

\section{CONCLUSIONS}

In this paper, we have explored and gleaned fascinating insights from spot price
data. We have seen the different dynamics that all play a role in affecting spot prices,
including location (both on a coarse-grained regional level and on a finer grained
availability zone level), type of instance, time of day, etc. We have analysed pricing
volatility, determining when's best to deploy an instance and whether we can
be confident in the reliability of the average price point shown (via the standard
deviation). We have also analysed modal price points for instance types in different AZs, giving us another viewpoint of spot price — and allowing us to make confident predictions on potential bid prices.

We have shown how important context is in interpreting these analyses and that taken
together, they paint a revealing picture about where best to deploy instances of varying
power for the lowest price.

The general conclusions we can therefore draw are:
\begin{enumerate}[I.]
\item The more powerful the instance, the cheaper it is in general to deploy in the Asia
Pacific or Canada regions - the less powerful the instance, the EU and US regions.
\item The Canada and Asia Pacific regions have a lot less demand, making it more
attractive for developers (both from a pricing point of view) but also as a means
of eliminating potential application ``hotspots".
\item AZs have a significant impact on price - it is possible to deploy very powerful
instances in a cheaper AZ located in an overall more expensive region. Figure \ref{fig:a4}
revealed to us how in the \emph{eu-west-1b} AZ, we could deploy a very powerful
\textit{r4.large} instance for a very low price (\$0.02 and \$0.12 an hour).
\item There are some exceptions to these general conclusions — one example is in
the US where \emph{m4.large} instances are amongst the most expensive to deploy
while not being that powerful an instance type.
\item Location plays a significant part in determining spot instance price, more than the
instance type itself.
\end{enumerate}
To conclude, spot Instances are very attractive as a means of using powerful compute power at very low prices. If a developer plans exactly where to deploy, examines all the pricing data
available, and takes all the analyses performed together and in context, it is entirely
possible to run workloads on Spot Instances at a very low risk of instance termination. Future work includes further analyses over longer periods of time, and building spot-price aware scheduling algorithms. 
\printbibliography
\balance
\end{document}